\documentclass[prd,twocolumn,preprintnumbers,nofootinbib,letter,superscriptaddress]{revtex4}

\usepackage{graphicx}
\usepackage{amssymb}
\usepackage{textcomp}
\usepackage{amsmath}
\usepackage{bm}
\usepackage{times}
\usepackage{epsfig}
\usepackage{color}
\usepackage{graphics}
\usepackage{hyperref}
\usepackage{setspace,ulem}
\allowdisplaybreaks

\hypersetup{
    pdfnewwindow=true,      
    colorlinks=true,       
    linkcolor=blue,          
    citecolor=blue,        
    filecolor=blue,      
    urlcolor=blue           
}

\usepackage{tikz}
\usetikzlibrary{trees}
\usetikzlibrary{decorations.pathmorphing}
\usetikzlibrary{decorations.markings}
\usetikzlibrary{decorations, decorations.markings, decorations.pathmorphing, arrows, graphs, shapes.geometric, snakes}
\usetikzlibrary{arrows}

\tikzset{
photon/.style={decorate, decoration={snake}, draw=red},
dark/.style={draw=gray, postaction={decorate},
        decoration={markings,mark=at position .55 with {\arrow[draw=gray]{>}}}},
antidark/.style={draw=gray, postaction={decorate},
        decoration={markings,mark=at position .55 with {\arrow[draw=gray]{<}}}},
electron/.style={draw=violet, postaction={decorate},
      decoration={markings,mark=at position .55 with {\arrow[draw=violet]{>}}}},
neutrino/.style={draw,color=violet,thick, postaction={decorate} },
quark/.style={draw=blue, postaction={decorate},
      decoration={markings,mark=at position .55 with {\arrow[draw=blue]{>}}}},
antiquark/.style={draw=blue, postaction={decorate},
    decoration={markings,mark=at position .55 with {\arrow[draw=blue]{<}}}},
gluon/.style={decorate, draw=or,
      decoration={coil,amplitude=2pt, segment length=3pt}},
ZZ/.style={decorate, decoration={snake,amplitude=1.5pt, segment length=5pt}, draw=greeen},
        left,
chargedscalar/.style={draw=black, postaction={decorate},
        decoration={markings,mark=at position .55 with {\arrow[scale=1.25,draw=black,>=latex]{>}}}},
 }

\definecolor{greeen}{rgb}{0.03,0.84,0.13}
\definecolor{test}{rgb}{0.03,0.74,0.33}
\definecolor{viol}{rgb}{0.44,0,0.94}
\definecolor{or}{rgb}{0.95,0.65,0}

\begin{document}

\preprint{ACFI T18-10}

\title{Doubly-Charged Scalars in the Type-II Seesaw Mechanism: \\
Fundamental Symmetry Tests and High-Energy Searches}

\author{P. S. Bhupal Dev}
\affiliation{Department of Physics and McDonnell Center for the Space Sciences, \\  Washington University, St. Louis, MO 63130, USA}

\author{Michael J. Ramsey-Musolf}
\affiliation{Amherst Center for Fundamental Interactions, Department of Physics, University of Massachusetts Amherst, Amherst, MA 01003, USA}
\affiliation{Kellogg Radiation Laboratory, California Institute of Technology, Pasadena, CA 91125, USA}

\author{Yongchao Zhang}
\affiliation{Department of Physics and McDonnell Center for the Space Sciences, \\  Washington University, St. Louis, MO 63130, USA}


\begin{abstract}
We analyze the sensitivity of low-energy fundamental symmetry tests to interactions mediated by doubly-charged scalars that arise in type-II seesaw models of neutrino mass and their left-right symmetric extensions. We focus on the next generation measurement of the parity-violating asymmetry in M{\o}ller scattering planned by the MOLLER collaboration at Jefferson Laboratory. We compare the MOLLER sensitivity to that of searches for charged lepton flavor violation (CLFV) and neutrinoless double beta-decay ($0\nu\beta\beta$-decay) as well as present and possible future high-energy collider probes. We show that for the simplest type-II seesaw scenario, CLFV searches have the greatest sensitivity. However, in a left-right symmetric extension where the scale of parity-breaking is decoupled from the $SU(2)_R$-breaking scale, the MOLLER experiment will provide a unique probe of scalar triplet interactions in the right-handed sector for a doubly-charged scalar mass up to $\sim$ 10 TeV and help elucidate the mechanism of $0\nu\beta\beta$-decay.
\end{abstract}
\maketitle


\section{Introduction}

Explaining the origin of the non-vanishing but tiny neutrino masses is a key open problem for particle physics. The simplest scenario entails introducing right-handed  neutrino (RHN) fields and Yukawa interactions akin to those for the charged elementary fermions of the Standard Model (SM). However, the required neutrino Yukawa couplings are significantly smaller (by at least six orders of magnitude) than the corresponding electron Yukawa coupling, $y_e$, a feature that many consider theoretically unappealing. An attractive alternative is the high-scale {type-I} seesaw mechanism~\cite{type1a, type1b, type1c, type1d, type1e}, wherein the RHNs carry a heavy Majorana mass $M_N$ -- the seesaw scale -- up to $\sim 10^{14}$ GeV. The corresponding Yukawa couplings can then be of order $\mathcal{O}(1)$, while the suppressed neutrino mass scale arises from the ratio of the weak and seesaw scales. {Several tree-level variations of this paradigm have been discussed over the years, such as type-II~\cite{Magg:1980ut, Schechter:1980gr, Mohapatra:1980yp, Lazarides:1980nt, Konetschny:1977bn, Cheng:1980qt}, III~\cite{Foot:1988aq}, inverse~\cite{Mohapatra:1986aw, Mohapatra:1986bd} and linear~\cite{Wyler:1982dd, Akhmedov:1995vm} seesaw models.}

If the seesaw mechanism is realized in nature, it is entirely possible that the seesaw scale $M_N$ is considerably lower than in the conventional picture. For example, if $M_N\sim 1$ TeV, the corresponding Yukawa couplings could be somewhat smaller than $y_e$ -- a situation that would not be wholly out of line compared to the vast spread in the magnitudes of the charged fermion Yukawa couplings. {Alternatively, larger Yukawa couplings could be made compatible with the neutrino oscillation data in a natural way in the inverse~\cite{Mohapatra:1986aw, Mohapatra:1986bd, Dev:2009aw} and linear~\cite{Wyler:1982dd, Akhmedov:1995vm, Malinsky:2005bi} seesaw models with small lepton number breaking. Yet another possibility is by making the vacuum expectation value (vev) responsible for neutrino mass generation much smaller than the electroweak scale, as in the case of type-II seesaw~\cite{Magg:1980ut, Schechter:1980gr, Mohapatra:1980yp, Lazarides:1980nt, Konetschny:1977bn, Cheng:1980qt}. In such low-scale seesaw scenarios, one could utilize laboratory experiments to probe the  predicted new particles and interactions.}

In this study, we consider the opportunity to exploit low-energy, high-precision experiments to probe the low-scale type-II seesaw mechanism. This genre of experiments -- sometimes denoted the {\it precision} or {\it sensitivity frontier} -- are  sensitive either to small deviations from the SM predictions or to rare phenomena that are highly suppressed or forbidden in the SM~\cite{RamseyMusolf:2009ga}. We focus in particular on the interplay of searches for charged lepton flavor violation (CLFV) and the neutrinoless double beta-decay ($0\nu\beta\beta$-decay) of heavy nuclei with a next generation measurement of the parity-violating asymmetry in fixed-target, polarized M{\o}ller scattering. While the sensitivity of CLFV and $0\nu\beta\beta$-decay for the type-II seesaw parameter space have been considered previously (see,  {\it e.g.}, Refs.~\cite{Cirigliano:2004tc, Abada:2007ux, Chakrabortty:2012mh, Awasthi:2015ota, Barry:2013xxa, Bambhaniya:2015ipg, Borah:2016iqd, Tello:2010am}), the opportunity with parity-violating M{\o}ller scattering has received less attention. Our study is motivated, in part, by the proposed MOLLER experiment~\cite{Benesch:2014bas, Moller} that is planned for the 12 GeV beam at Jefferson Lab. In the absence of physics beyond the SM (BSM), the MOLLER asymmetry measurement will determine the scale evolution of the weak mixing angle to unprecedented precision. Any deviation from the SM expectation could signal the presence of BSM scenarios, such as a heavy $Z^\prime$ gauge boson~\cite{Langacker:2008yv, Li:2009xh} (not much room left for this possibility), a light \lq\lq dark Z" boson~\cite{Davoudiasl:2012ag,Davoudiasl:2012qa}, or  $R$-parity conserving and violating supersymmetric models~\cite{Kurylov:2003zh, RamseyMusolf:2006vr}.

Our emphasis in this work falls on the possible signatures of the scalar {isospin}-triplets in the type-II seesaw~\cite{Magg:1980ut, Schechter:1980gr, Mohapatra:1980yp, Lazarides:1980nt, Konetschny:1977bn, Cheng:1980qt} and its left-right symmetric extensions~\cite{LR1, LR2, LR3} as a case study. These triplets and their interactions with the SM leptons are a key ingredient in the type-II scenario:
\begin{itemize}
\item
The simplest type-II scenario involves the scalar multiplet $\Delta_L$ that transforms as a complex triplet under the SM $SU(2)_L$ gauge group. The neutral component of $\Delta_L$ receives a non-zero vev $v_L/\sqrt{2}$, leading to the light neutrino Majorana mass matrix
\begin{equation}
m_\nu = \sqrt 2 f_L v_L \, ,
\label{eq:typeIImass}
\end{equation}
where $f_L$ is a $3\times 3$ matrix of triplet Yukawa couplings to the left-handed (LH) lepton-doublet fields. The LH doubly-charged scalar component $H_L^{\pm\pm}$ of the triplet $\Delta_L$ couples to same-sign charged leptons, with the Yukawa couplings $f_L$ intimately related to the neutrino mass and mixing data~\cite{PDG}. The electron element, i.e. $(f_L)_{ee}$, would mediate the low-energy electron-electron scattering, and thus get constrained by the upcoming MOLLER data, as shown in Section~\ref{sec:moller}.\footnote{{These results also apply to leptophilic doubly-charged scalars appearing in other neutrino mass models, such as the Georgi-Machacek model~\cite{Georgi:1985nv,  Chanowitz:1985ug}, Zee-Babu model~\cite{Zee:1985id, Babu:1988ki} and Babu-Nandi-Tavartkiladze model~\cite{Babu:2009aq}.} }
\item
The left-right symmetric model (LRSM)~\cite{LR1, LR2, LR3}, originally proposed as a minimal extension of the SM for providing an alternative approach to parity violation in low-energy processes, has emerged as a well-motivated model for neutrino masses via the type-I~\cite{type1a, type1b, type1c, type1d, type1e} and/or type-II~\cite{Magg:1980ut, Schechter:1980gr, Mohapatra:1980yp, Lazarides:1980nt, Konetschny:1977bn, Cheng:1980qt} seesaw mechanisms. The LRSM has all the important ingredients of type-II seesaw and thus turns out to be also a natural extension of the minimal type-II seesaw. In addition, the extra scalar and gauge bosons and RHNs in the heavy right-handed (RH) $SU(2)_R$ sector are also crucial for the neutrino mass physics~\cite{Deppisch:2015qwa} and might also be tested in the high-intensity/precision frontier~\cite{Castillo-Felisola:2015bha, Dev:2016vle, Dev:2017dui, Cottin:2018kmq, Nemevsek:2018bbt}, including the proposed SHiP~\cite{Alekhin:2015byh} and MATHUSLA~\cite{Curtin:2018mvb} experiments. As a ``partner'' of $H_L^{\pm\pm}$ under parity, there exists an RH doubly-charged scalar $H_R^{\pm\pm}$, originating from an  $SU(2)_R$-triplet scalar $\Delta_R$ and coupling to the RH charged leptons via the Yukawa coupling matrix $f_R$. Parity symmetry requires that the (gauge and) Yukawa couplings of $H_{L,\,R}^{\pm\pm}$ are the same, i.e. $f_L = f_R$, which has profound implications for the light and heavy neutrinos, as well as for the low energy M{\o}ller scattering, as shown in Section~\ref{sec:lrsm}.
\end{itemize}

With the neutrino mass formula in Eq.~(\ref{eq:typeIImass}),
all the elements of $f_L$ are correlated by neutrino oscillation data. These elements also include the flavor non-diagonal couplings relevant to CLFV processes, such as $\mu \to eee$ and $\mu \to e\gamma$. Given the correlation of all couplings through neutrino oscillation phenomenology, the element $(f_{L})_{ee}$ is, thus,  also subject to the stringent CLFV limits (see, {\it e.g.}, Ref.~\cite{Dev:2017ouk}). In the LRSM with parity symmetry, one has $f_L=f_R$, so the same constraints will apply in this scenario as well.
As we show in Sections~\ref{sec:left} and~\ref{sec:lrsm}, even if the neutrino data uncertainties are taken into consideration, the MOLLER sensitivity is superseded by CLFV bounds, independent of the present lower bounds on the $H_L^{\pm\pm}$ obtained from direct collider searches (see below for a full discussion). Consequently,  if the MOLLER experiment yields a deviation from the SM, one would need to extend the pure type-II seesaw or the parity-conserving LRSM in a manner consistent with the CLFV constraints.

Indeed, if parity symmetry is not completely restored at the TeV scale in the LRSM, then the CLFV constraints on the $(f_{R})_{ee}$ no longer apply. Theoretically, this possibility has been considered previously.
Some of the Yukawa couplings could be different ($f_L \neq f_R$), for instance, in the LRSM with $D$-parity breaking~\cite{CMP} where by introducing a parity-odd singlet with high-scale vev, one can give a large mass to $\Delta_L$ so that it decouples from the low-energy theory. Experimentally, the MOLLER experiment could probe a large parameter space of this scenario that is beyond the reach of past, current and future low- and high-energy experiments, such as
\begin{itemize}
\item direct same-sign dilepton pair searches at $\sqrt s=13$ TeV LHC~\cite{Aaboud:2017qph, CMS:2017pet}, which are roughly 800 GeV and 657 GeV for the LH and RH doubly-charged scalars, respectively;
\item the lower-energy constraints on $H_R^{\pm\pm}$ from the LEP $e^+ e^- \to e^+ e^-$ (Bhabha scattering) data~\cite{Abdallah:2005ph};
\item the non-observation $0\nu\beta\beta$-decays in EXO-200~\cite{Albert:2017owj}, KamLAND-Zen~\cite{KamLAND-Zen:2016pfg}, GERDA~\cite{Agostini:2018tnm}, MAJORANA DEMONSTRATOR~\cite{Aalseth:2017btx}, CUORE~\cite{Alduino:2017ehq} and NEMO-3~\cite{Arnold:2018tmo}, and the prospect in the ongoing and upcoming $0\nu\beta\beta$ experiments~\cite{Agostini:2017jim}.
\end{itemize}
{Only a direct measurement of the Yukawa coupling $(f_R)_{ee}$ at future lepton colliders, such as CEPC~\cite{CEPC}, FCC-ee~\cite{Gomez-Ceballos:2013zzn}, ILC~\cite{Baer:2013cma} or CLIC~\cite{Battaglia:2004mw} could surpass the MOLLER sensitivity for the entire parameter space of interest (see Fig.~\ref{fig:final2}).}

The remaining sections are organized as follows: The MOLLER prospects are sketched in Section~\ref{sec:moller}. Section~\ref{sec:left} is devoted to the LH doubly-charged scalar in the minimal type-II seesaw model. Section~\ref{sec:lrsm} focuses on the RH doubly-charged scalar in parity-conserving LRSM. The parity-violating case of LRSM follows in Section~\ref{sec:lrsm2}. We conclude in Section~\ref{sec:conclusion}.


\section{MOLLER prospects}
\label{sec:moller}

\begin{figure}[t!]
\includegraphics[width=3.5cm]{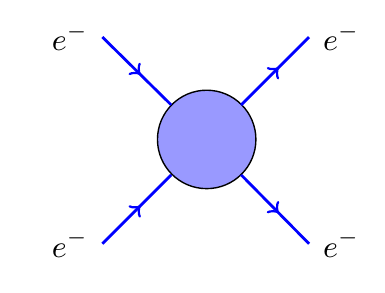} \hspace{0.2cm}
\includegraphics[width=4cm]{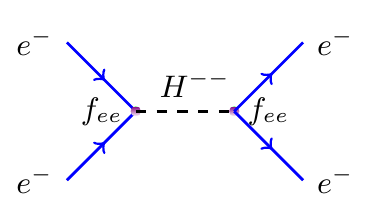}
\caption{Effective 4-fermion interaction (left) and doubly-charged scalar contribution (right) for the M{\o}ller scattering. }\label{fig:feyn}
\end{figure}

The MOLLER collaboration proposes to measure the parity-violating asymmetry $A_{\rm PV}$ in the scattering of longitudinally polarized electrons off unpolarized electrons at Jefferson Lab to an overall precision of 0.7 ppb, which will measure the weak charge of the electron $Q_W^e$ to an overall fractional accuracy of 2.4\%~\cite{Benesch:2014bas, Moller}.\footnote{It has also been proposed to measure the weak mixing angle in electron-proton scattering experiment P2~\cite{Becker:2018ggl}, with a comparable sensitivity to MOLLER, which is however not relevant to the doubly-charged scalars, which are hadrophobic and do not couple directly to quarks.}  This gives a model-independent sensitivity to new four-electron contact interaction (Fig.~\ref{fig:feyn} left panel) amplitude as
\begin{align}
\frac{\Lambda}{\sqrt{|g_{RR}^2-g_{LL}^2|}} \ = \ \frac{1}{\sqrt{\sqrt{2}G_F|\Delta Q_W^e|}} \ \simeq \ 7.5~{\rm TeV} \, ,
\label{eq:4fermi}
\end{align}
where $g_{LL,RR}$ are the coupling constants for the new vector and axial vector interactions between LH and RH electrons, respectively, and $G_F$ is the Fermi coupling constant. Here, we have used $|\Delta Q_W^e/Q_W^e|=0.024$, with $Q_W^e= - 1 + 4\sin^2\theta_W$ at tree level, $\theta_W$ being the weak mixing angle. {We also take into account the impact of one-loop electroweak radiative corrections, which reduce the magnitude of $Q_W^e$ by $\sim$ 40\% compared to the nominal tree-level value~\cite{Benesch:2014bas}.}

Since the LH and RH doubly-charged scalars couple to two electrons [cf.~Eqs.~(\ref{eqn:lagrangian}) and (\ref{eqn:lagrangian2}) below], the corresponding $s$-channel exchange four-electron amplitude for M{\o}ller scattering $e^-e^-\to H_{L,R}^{--}\to e^-e^-$ (Fig.~\ref{fig:feyn} right panel) that violates lepton number by two units at each vertex can be written as
\begin{align}
{\cal M}_{\rm PV} \ \sim \ \frac{|(f_{L})_{ee}|^2}{2 M_{H_L^{\pm\pm}}^2}(\bar{e}_L\gamma^\mu e_L)(\bar{e}_L\gamma_\mu e_L) ~+~ (L \leftrightarrow R) \,.
\end{align}
If we just keep the $H_L^{\pm\pm}$, this is equivalent to a contact four-fermion interaction with the effective cutoff scale $\Lambda=M_{L}^{\pm\pm}$ with $|g_{LL}|^2=|(f_{L})_{ee}|^2/2$ and $g_{RR}=0$ in Eq.~\eqref{eq:4fermi}. The agreement between the proposed $A_{\rm PV}$ measurement and the SM prediction would therefore constrain the ratio of the LH doubly-charged scalar mass $M_{H_L^{\pm\pm}}$ and the Yukawa coupling $(f_L)_{ee}$ to electrons at the level of
\begin{align}
\label{eqn:moller}
\frac{M_{H_{L}^{\pm\pm}}}{|(f_L)_{ee}|} \ \gtrsim \ 3.7~{\rm TeV} 
\,,
\end{align}
at 95\% confidence level (CL),\footnote{Note that the reach of 5.3 TeV reported in Ref.~\cite{Benesch:2014bas} is at the $1\sigma$ level.} which applies equally to the RH doubly-charged scalar (with $L\leftrightarrow R$ in the equation above).
This sensitivity does not depend on how the doubly-charged scalars $H_{L,R}^{\pm\pm}$ decay or how they couple to other (beyond) SM particles, and it is largely complementary to the direct searches of LH and RH doubly-charged scalars at high-energy colliders. {We emphasize in particular that depending on the magnitude of $f_{ee}$ ($L,R$ subscripts suppressed), the  mass reach in Eq.~(\ref{eqn:moller}) could exceed  the {prospective high-luminosity LHC reach and even the future 100 TeV $pp$ collider} reach~\cite{Arkani-Hamed:2015vfh, Dev:2016dja, Contino:2016spe}. Of course, the scale of $|f_{ee}|$ will depend on the specific type-II seesaw implementation and the corresponding value(s) of the triplet vev(s).}


\begin{figure*}[!t]
  \centering
  \includegraphics[height=0.25\textwidth]{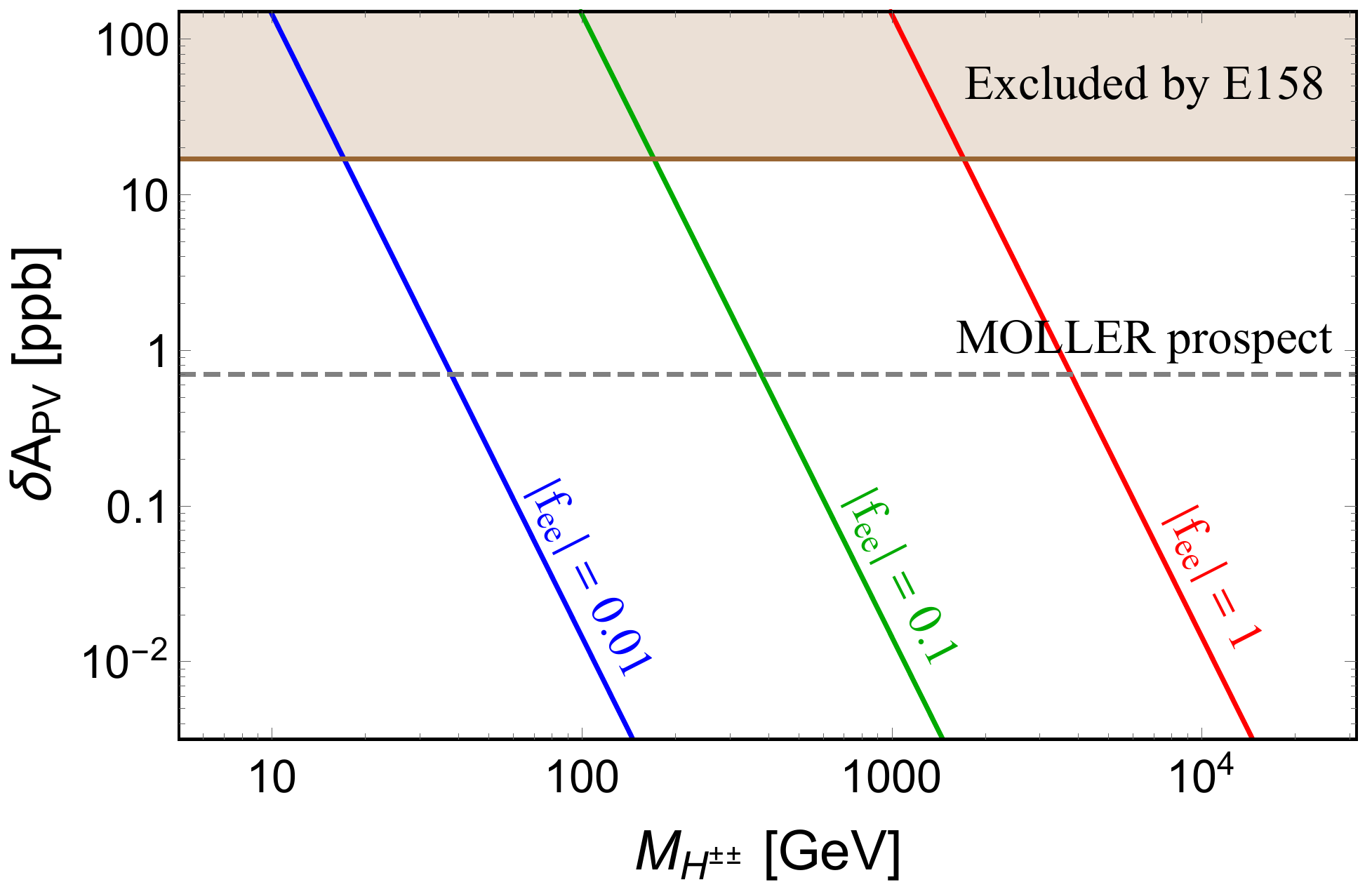}
\hspace{0.2cm}
  \includegraphics[height=0.25\textwidth]{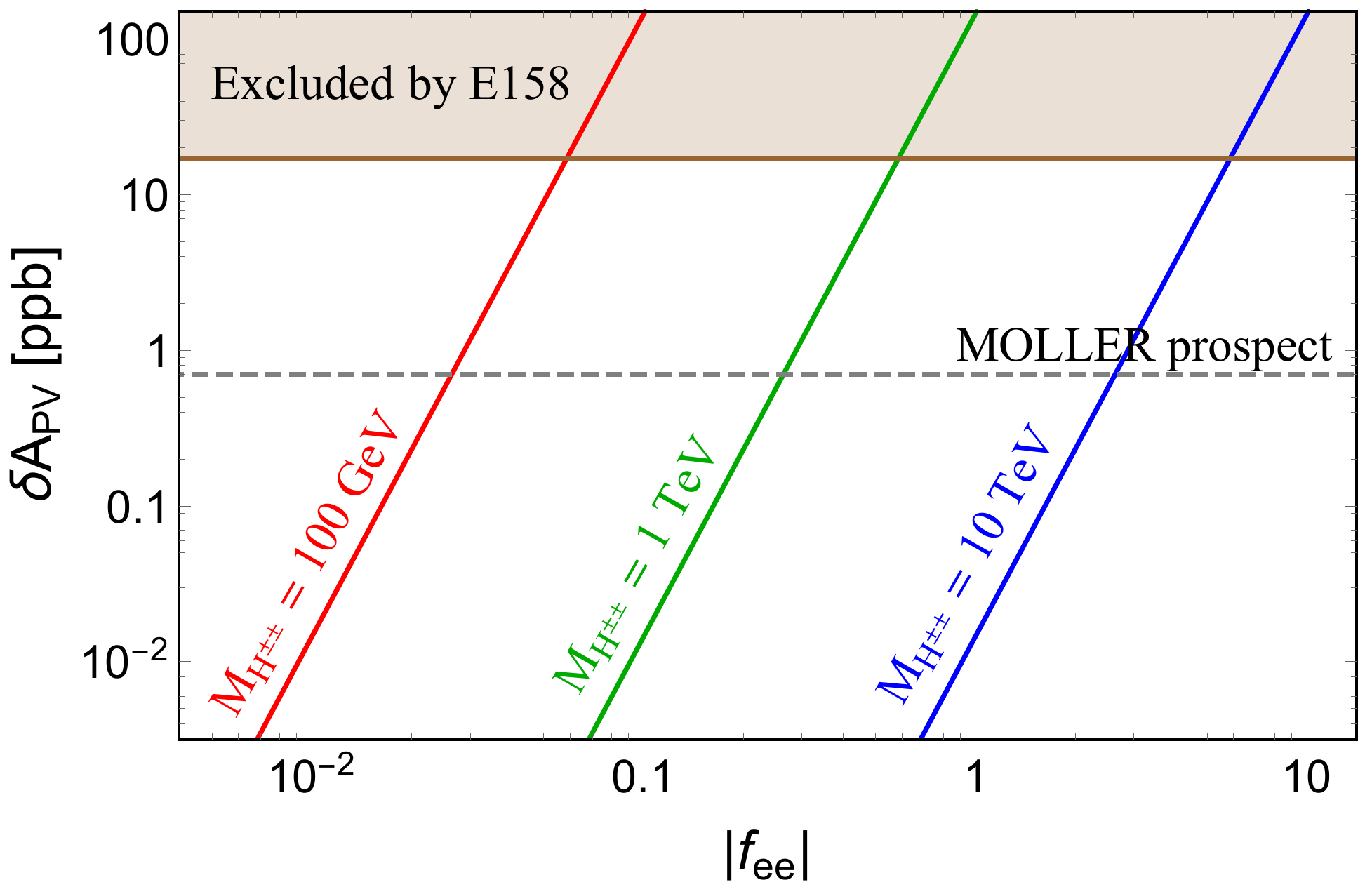}
  \caption{{\it Left:} The contribution of $H^{\pm\pm}$ (either LH or RH) to the parity-violating asymmetry $\delta A_{\rm PV}$ in the MOLLER experiment, as a function of the doubly-charged scalar mass $M_{H^{\pm\pm}}$, for three benchmark values of the Yukawa coupling $|f_{ee}| = 0.01$ (blue), 0.1 (green) and $1$ (red). {\it Right:} $\delta A_{\rm PV}$ as a function of the coupling $|f_{ee}|$ for three representative values of the doubly-charged scalar mass $M_{H^{\pm\pm}} = 100$ GeV (red), 1 TeV (green) and 10 TeV (blue). In both panels, the shaded region at the top is excluded by the E158 experiment~\cite{Anthony:2005pm}, and the horizontal dashed line indicates the projected MOLLER sensitivity~\cite{Benesch:2014bas}.}
  \label{fig:prospect}
\end{figure*}

To obtain additional intuition for the interplay of $f_{ee}$,  $M_{H^{\pm\pm}}$, and the MOLLER sensitivity in Eq.~(\ref{eqn:moller}), we show in the left panel of Fig.~\ref{fig:prospect} the contribution of $H^{\pm\pm}$ (either LH or RH) to the parity-violating asymmetry in the MOLLER experiment, dubbed as $\delta A_{\rm PV}$, as a  function of the doubly-charged scalar mass {$M_{H^{\pm\pm}}$} for three benchmark values of {$|f_{ee}| = 0.01$} (blue), 0.1 (green) and 1 (red). In the right panel of Fig.~\ref{fig:prospect}, the $\delta A_{\rm PV}$ is depicted as a function of the Yukawa coupling {$|f_{ee}|$} for three benchmark masses of {$M_{H^{\pm\pm}} = 100$} GeV (red), 1 TeV (green) and 10 TeV (blue). Note that in the simplest type-II see saw scenario, the scale of $v_L$ goes roughly as $m_\nu/ f_L$ [cf. Eq.~\eqref{eq:typeIImass}]. Electroweak precision tests require that $v_L\lesssim 5$ GeV~\cite{delAguila:2008ks, Dev:2013ff}. The ranges for $f_{ee}$ indicated in Fig.~\ref{fig:prospect} {are consistent with these constraints, given that $m_\nu\lesssim$ eV.} In both  panels of Fig.~\ref{fig:prospect}, the shaded regions with $\delta A_{\rm PV} > 17$ ppb are excluded by the current most stringent limit from E158~\cite{Anthony:2005pm}. The MOLLER experiment could reach a higher precision of 0.7 ppb~\cite{Benesch:2014bas}, as indicated by the horizontal dashed line in Fig.~\ref{fig:prospect}, which would probe a {$H_{L,R}^{\pm\pm}$} mass up to $\simeq 10$ TeV, {as long as $f_{ee}$ remains perturbative.} Looking ahead, we also note that the illustrative sensitivities in Fig.~\ref{fig:prospect} will be most relevant to the LRSM with parity-violation (Section~\ref{sec:lrsm2}), as the bounds from CLFV searches supersede that of the MOLLER sensitivity for the minimal type-II and parity-conserving LRSM, as shown below.


\section{Left-handed doubly-charged scalar in type-II seesaw}
\label{sec:left}

In the minimal type-II seesaw, in addition to the SM Higgs doublet $\phi = \left( \phi^+ ,\, \phi^0 \right)^{\sf T}$, one introduces a complex $SU(2)_L$ scalar triplet that can be written as
\begin{align}
\label{eqn:DeltaL}
\Delta_L \ = \  \left(\begin{array}{cc}
\delta_L^+/\sqrt{2} & \delta_L^{++} \\
\delta_L^0 & -\delta_L^+/\sqrt{2}
\end{array}\right) \, .
\end{align}
A non-zero vev for the Higgs doublet field $\langle \phi^0 \rangle = v_{\rm EW}/\sqrt2$ (with $v_{\rm EW} \simeq$ 246 GeV being the electroweak scale) induces a tadpole term for the scalar triplet field ${\Delta}_L$, thereby generating a non-zero vev for its neutral component, $\langle \delta_L^0\rangle = v_L/\sqrt 2$, and breaking lepton number by two units in the presence of the interaction (\ref{eqn:lagrangian}) below.
As the vev $v_L$ sets the scale for the light neutrino masses, it is expected to be much smaller than the electroweak scale, possibly even close to the eV scale. As noted above, electroweak precision data require that $v_L\lesssim 5$ GeV~\cite{delAguila:2008ks, Dev:2013ff}. In the limit of $v_L \ll v_{\rm EW}$, after spontaneous symmetry breaking, we obtain the neutral CP-even component $H \simeq {{\rm Re} \, \delta_L^0}/{\sqrt2}$, the CP-odd component $A \simeq {{\rm Im} \, \delta_L^0}/{\sqrt2}$, the singly-charged scalar $H^{\pm} \simeq \delta_L^\pm$, and the doubly-charged scalar $H_L^{\pm\pm} = \delta_L^{\pm\pm}$.

The triplet $\Delta_L$ couples to the SM lepton doublet  $\psi_L =(\nu,\ell)_L^{\sf T}$ via the Yukawa interactions
\begin{eqnarray}
\label{eqn:lagrangian}
{\cal L}_Y \ = \
- \left(f_L \right)_{\alpha\beta} \psi_{L,\,a\alpha}^{\sf T}C \varepsilon_{ab} {\Delta}_L \psi_{L,\,b\beta} ~+~ {\rm H.c.},
\end{eqnarray}
where $a,\,b$ are the isospin indices, $\alpha,\,\beta=e,\mu,\tau$ denote the lepton flavor, $C$ is the charge conjugation matrix and {$\varepsilon_{ab}$ the antisymmetric tensor}. Then the light neutrino mass matrix is obtained with the induced vev $v_L$ [cf.~Eq.~\eqref{eq:typeIImass}]:
\begin{eqnarray}
\label{eqn:neutrino}
m_\nu \ = \
\sqrt2 \, f_L v_L \ = \
U \widehat{m}_\nu U^{\sf T} \, ,
\end{eqnarray}
where $\widehat{m}_\nu = {\rm diag} \{ m_1,\, m_2,\, m_3 \}$ the diagonal neutrino masses and $U$ is the standard PMNS mixing matrix.
Thus, the Yukawa coupling matrix $f_L$ is fixed by neutrino oscillation data~\cite{PDG}: the observed neutrino mass squared differences and mixing angles, up to the unknown lightest neutrino mass scale $m_0$, the neutrino mass hierarchy, and the CP violating phases.  


\subsection{Constraints}

{For phenomenological purposes, it is reasonable to assume that the triplet scalars are mass degenerate at the tree-level; in this case the mass splitting $M_{H_L^{\pm\pm}} - M_{H^\pm} \simeq 540$ MeV can be induced at the one-loop level by interactions  with the SM gauge bosons~\cite{Cirelli:2005uq}. Then the decay $H_L^{\pm\pm} \to H^\pm W^{\pm \ast}$ is expected to be highly suppressed. For sufficiently small $v_L$, the coupling of $H_L^{\pm\pm}$ to a same sign $W$ boson pair is also highly suppressed. One finds that for $v_L \lesssim 0.1$ MeV, the LH doubly-charged scalar $H_L^{\pm\pm}$ decays predominantly into a same-sign dilepton pair}, {\it i.e.} $H_L^{\pm\pm} \to \ell_\alpha^\pm \ell_\beta^\pm$~\cite{Perez:2008ha, Melfo:2011nx, Kanemura:2014goa}. At high energy colliders, such processes are almost background free, and the most stringent mass limits on $H_L^{\pm\pm}$ are obtained from direct dilepton searches at the $\sqrt s=13$ TeV LHC~\cite{Aaboud:2017qph, CMS:2017pet}. From the Drell-Yan production $pp \to \gamma^\ast/Z^\ast \to H_L^{++} H_L^{--}$ and the subsequent decays $H_L^{\pm\pm} \to e^\pm e^\pm$, 
 the current LHC limit is roughly $M_{H_L^{\pm\pm}} \gtrsim 600-800$ GeV, depending on the branching fraction of $H_L^{\pm\pm}$ to the di-electron channel. The limits are expected to be more constraining if the  photon fusion process $\gamma\gamma \to H_L^{++} H_L^{--}$ is also taken into consideration~\cite{Babu:2016rcr}. With more data from 13 TeV LHC and future 14 TeV and high-luminosity stages, the doubly-charged scalar could be probed up to a TeV or so. Future 100 TeV hadron colliders like SPPC~\cite{Tang:2015qga} or FCC-hh~\cite{fcc-hh} would push the reaches even higher, but as far as we know, there is no dedicated study on the future prospects of $H_L^{\pm\pm}$ at 100 TeV collider.

Given the LFV couplings $(f_L)_{\alpha\beta}$, the doubly-charged scalar $H_L^{\pm\pm}$ could induce rare flavor violating decays such as $\ell_\alpha \to \ell_\beta \ell_\gamma \ell_\delta$, $\ell_\alpha \to \ell_\beta \gamma$, as well as contribute to the anomalous magnetic moments of electron and muon, and muonium-anti-muonium oscillation~\cite{Dev:2018upe} which are all highly suppressed in the SM~\cite{PDG}.
Among the limits obtained from studies of these processes,  the most stringent are those from $\mu \to eee$ and $\mu \to e\gamma$. The partial widths for these tree and loop level processes are respectively~\cite{Barry:2013xxa, Akeroyd:2009nu, Dinh:2012bp, Mohapatra:1992uu, Leontaris:1985qc, Swartz:1989qz, Cirigliano:2004mv, Pal:1983bf}
\begin{eqnarray}
\label{eqn:mu2eee}
{\rm BR} (\mu \to eee) & \ \simeq \ &
\frac{|(f_L)_{ee}^\dagger (f_L)_{e\mu}|^2}{4 G_F^2 M_{H_L^{\pm\pm}}^4} \,, \\
\label{eqn:mu2egamma}
{\rm BR} (\mu \to e \gamma) & \ \simeq \ &
\frac{\alpha_{\rm EM} |\sum_\ell (f_L)_{\mu\ell}^\dagger (f_L)_{e\ell}|^2}{3\pi G_F^2 M_{H_L^{\pm\pm}}^4} \,,
\end{eqnarray}
where $\alpha_{\rm EM}$ is the fine structure constant and in Eq.~(\ref{eqn:mu2egamma}) we have summed up all the diagrams involving $\ell = e,\, \mu,\, \tau$ lepton running in the loop.
The current limits of ${\rm BR} (\mu \to eee) < 1.0 \times 10^{-12}$~\cite{Bellgardt:1987du} and ${\rm BR} (\mu \to e \gamma) < 4.2 \times 10^{-13}$~\cite{MEG:2016wtm} put severe constraints on the combinations of LFV couplings in Eqs.~(\ref{eqn:mu2eee}) and (\ref{eqn:mu2egamma}), which correspond to an effective cut-off scale of $\Lambda \simeq M_{H_L^{\pm\pm}} / \sqrt{|f^\dagger f|}$:
\begin{eqnarray}
\mu \to eee : && \;\;\;
\frac{M_{H_L^{\pm\pm}}}{\sqrt{|(f_L)_{ee}^\dagger (f_L)_{e\mu}|}} > 208 \, {\rm TeV} \,, \\
\mu \to e\gamma : && \;\;\;
\frac{M_{H_L^{\pm\pm}}}{\sqrt{|\sum_\ell (f_L)_{\mu\ell}^\dagger (f_L)_{e\ell}|}} > 61 \, {\rm TeV} \,.
\end{eqnarray}
{These limits are clearly more stringent than the MOLLER prospects given by Eq.~\eqref{eqn:moller}; see below for more details.}

\subsection{MOLLER prospect}

\begin{table}[!t]
  \centering
  \caption[]{Best-fit values and $2\sigma$ ranges~\cite{PDG} of the neutrino mass square differences and mixing angles for NH and IH used in our numerical analysis.   The Dirac CP phase $\delta_{\rm CP}$ and the Majorana phases $\alpha$ and $\beta$ are left unconstrained. 
}
  \label{tab:neutriodata}
  \begin{tabular}[t]{ccc}
  \hline\hline
  parameters & NH & IH \\ \hline
  $\Delta m^2_{21}$ [eV$^2$] & $(7.53 \pm 0.18) \times 10^{-5}$ & $(7.53 \pm 0.18) \times 10^{-5}$ \\
  $|\Delta m^2_{32}|$ [eV$^2$] & $(2.45 \pm 0.05) \times 10^{-3}$ & $(2.52 \pm 0.05) \times 10^{-3}$ \\ \hline

  $\sin^2\theta_{12}$ & $0.307 \pm 0.013$ & $0.307 \pm 0.013$ \\
  $\sin^2\theta_{23}$ & $0.51 \pm 0.04$ & $0.50 \pm 0.04$ \\
  $\sin^2\theta_{12}$ & $0.021 \pm 0.0011$ & $0.021 \pm 0.0011$ \\ \hline
  $\delta_{\rm CP}$ & $[0,\, 2\pi]$ & $[0,\, 2\pi]$ \\
  $\alpha$ & $[0,\, 2\pi]$ & $[0,\, 2\pi]$ \\
  $\beta$ & $[0,\, 2\pi]$ & $[0,\, 2\pi]$ \\
  \hline\hline
  \end{tabular}
\end{table}

\begin{figure*}[t!]
  \centering
  \includegraphics[height=0.25\textwidth]{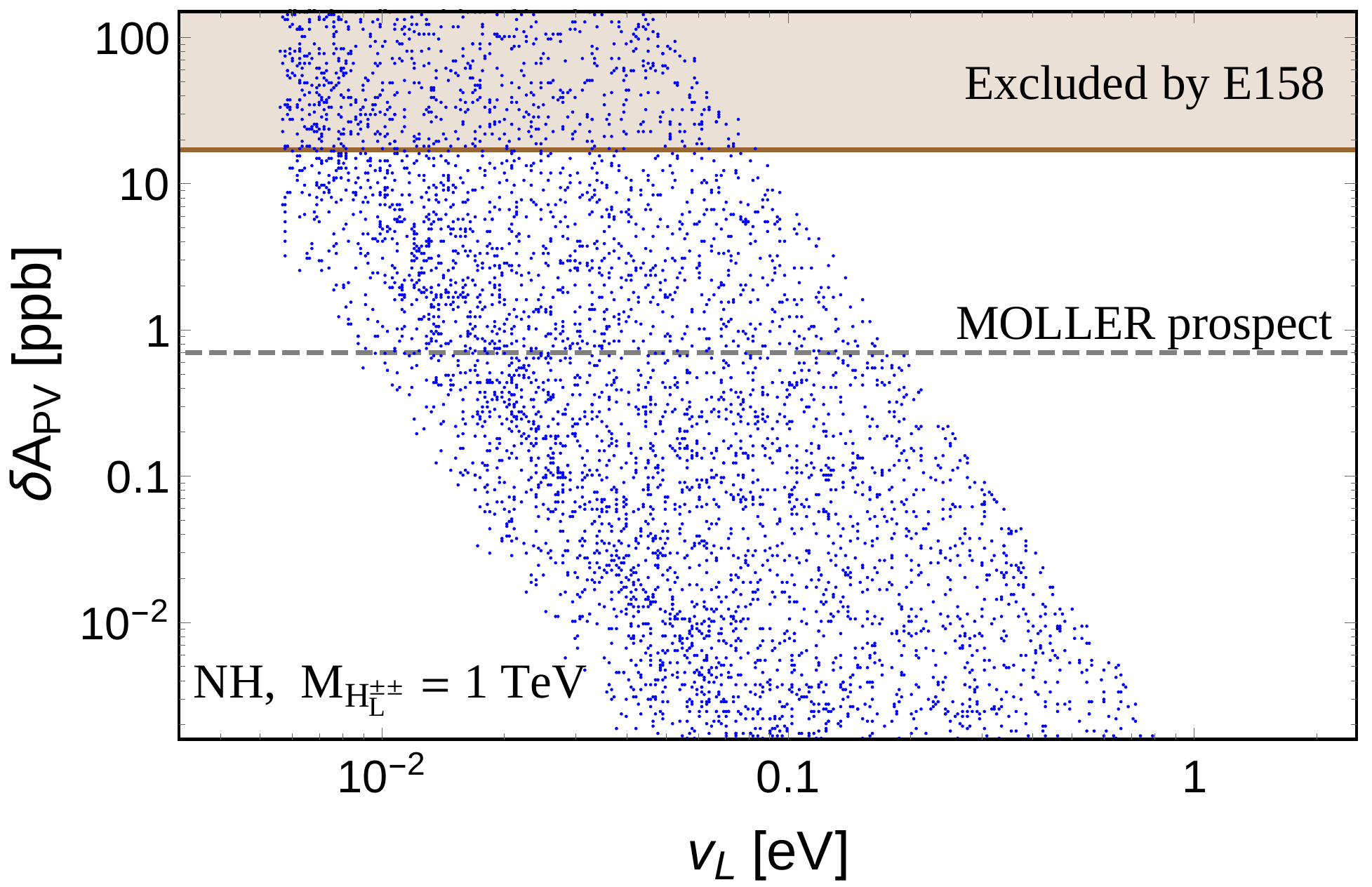}
\hspace{0.2cm}
  \includegraphics[height=0.25\textwidth]{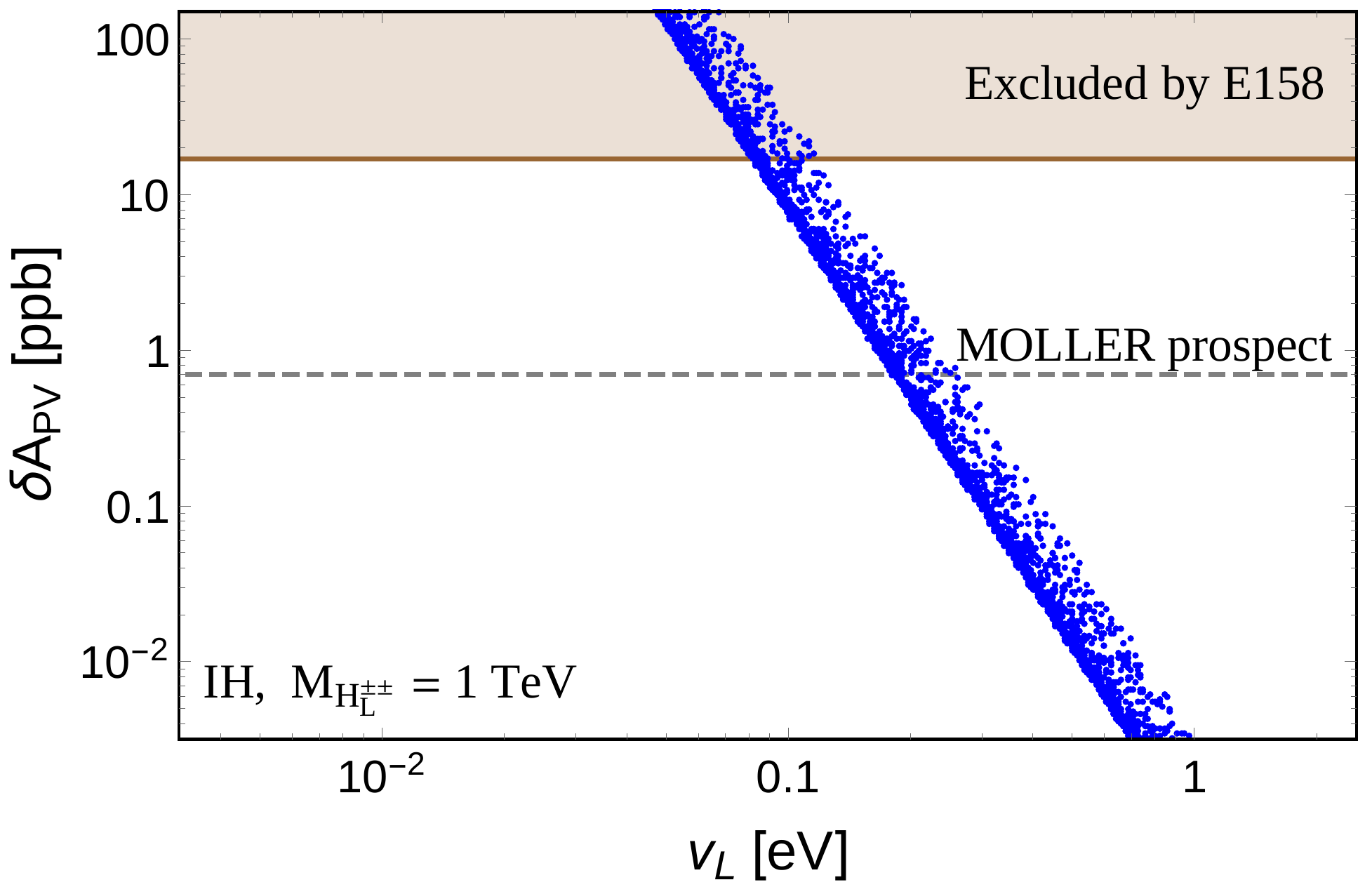}
  \caption{The contribution of $H_L^{\pm\pm}$ to the parity-violating asymmetry $\delta A_{\rm PV}$ in the MOLLER experiment, as a function of the vev $v_L$ in the minimal type-II seesaw for the doubly-charged scalar mass $M_{H_L^{\pm\pm}} = 1$ TeV. The left (right) plot is for NH (IH). In both panels, the shaded region is excluded by E158~\cite{Anthony:2005pm}, and the horizontal line indicates the projected MOLLER sensitivity~\cite{Benesch:2014bas}.}
  \label{fig:left1}
\end{figure*}

In light of the type-II seesaw relation in Eq.~(\ref{eqn:neutrino}), the projected MOLLER limit on the Yukawa coupling $(f_L)_{ee}$ in Fig.~\ref{fig:prospect} can be applied to the triplet vev $v_L$ via $v_L  = (m_\nu)_{ee} / \sqrt2 (f_L)_{ee}$. Here $(m_\nu)_{ee}$ is nothing but the effective electron neutrino mass in $0\nu\beta\beta$ decays. However, in the pure type-II seesaw, the contribution of $H_L^{\pm\pm}$ to $0\nu\beta\beta$ decay is suppressed by the doubly-charged scalar mass compared to the canonical terms induced by the Majorana neutrino mass $(m_\nu)_{ee}$ [cf.~Eq.~\eqref{eqn:mnuee}]. Thus in the minimal type-II seesaw, the $0\nu\beta\beta$ can not set any limits on the doubly-charged scalar $H_L^{\pm\pm}$~\cite{Chakrabortty:2012mh}. We adopt the neutrino mass and mixing data from Ref.~\cite{PDG} for both normal hierarchy (NH) and inverted hierarchy (IH), which are collected in Table~\ref{tab:neutriodata}. {Though the recent T2K~\cite{Abe:2017vif} and NO$\nu$A~\cite{NOvA:2018gge} results indicate a preference for non-zero $\delta_{\rm CP}$, this has not been established at $5\sigma$ level; therefore, we vary it within the whole range of $[0,\, 2\pi]$.}

To take into consideration the uncertainties of neutrino data, we vary the solar and atmospheric neutrino mass squared differences and the three mixing angles within their $2\sigma$ ranges, as shown in Table~\ref{tab:neutriodata}.  The lightest neutrino mass $m_0$ is unconstrained by the oscillation data, and we vary it between $[0,\, 0.05 \, {\rm eV}]$ to satisfy the cosmological limit on the sum of neutrino masses $\sum_i m_{\nu,i} < 0.23$ eV~\cite{Ade:2015xua}.
The value of $v_L$ is taken to be from $10^{-3}$ eV to 10 eV.\footnote{If $v_L$ is too large, the coupling $(f_L)_{ee}$ will be very small and out of the MOLLER reach; see Fig.~\ref{fig:left1}.} We assume all the input parameters obtained from neutrino oscillation data, the lightest neutrino mass and the vev $v_L$ are uniformly distributed in their corresponding ranges. Then we obtain the coupling $(f_L)_{ee}$ from these input parameters by using Eq.~(\ref{eqn:neutrino}) and compare it to the MOLLER sensitivity.

Fig.~\ref{fig:left1} gives the resulting scatter plots for the shift in the parity-violating asymmetry $\delta A_{\rm PV}$ as a function of $v_L$, for both NH (left) and IH (right) of neutrino masses, with the doubly-charged scalar mass fixed at $M_{H_L^{\pm\pm}} = 1$ TeV. As in Fig.~\ref{fig:prospect}, the shaded region is excluded by E158~\cite{Anthony:2005pm} and the horizontal dashed line gives the MOLLER sensitivity~\cite{Benesch:2014bas}. One can see from these plots that the MOLLER experiment is sensitive to a small vev
\begin{eqnarray}
v_L \ \lesssim \ 0.3 \, {\rm eV} \times
\left( \frac{M_{H_L^{\pm\pm}}}{1 \, {\rm TeV}} \right)^{-1} \,.
\end{eqnarray}
{For heavier (lighter) doubly-charged scalar $H_L^{\pm\pm}$, the Yukawa coupling $(f_L)_{ee}$ is expected to be larger (smaller), and the MOLLER experiment is sensitive to a smaller (larger) $v_L$ in Fig.~\ref{fig:left1}.} In the NH case, the element $(m_\nu)_{ee}$ is rather sensitive to the lightest neutrino mass $m_1$, thus the resultant scattered band is rather broad in the left panel of Fig.~\ref{fig:left1}. In the IH case, $(m_\nu)_{ee}$ is almost independent of the lightest neutrino mass $m_3$, thus the band in the right panel of Fig.~\ref{fig:left1} is much narrower, which reflects only the uncertainties of the mass squared differences and mixing angles in Table~\ref{tab:neutriodata}.

\begin{figure*}[t!]
  \centering
  \includegraphics[height=0.32\textwidth]{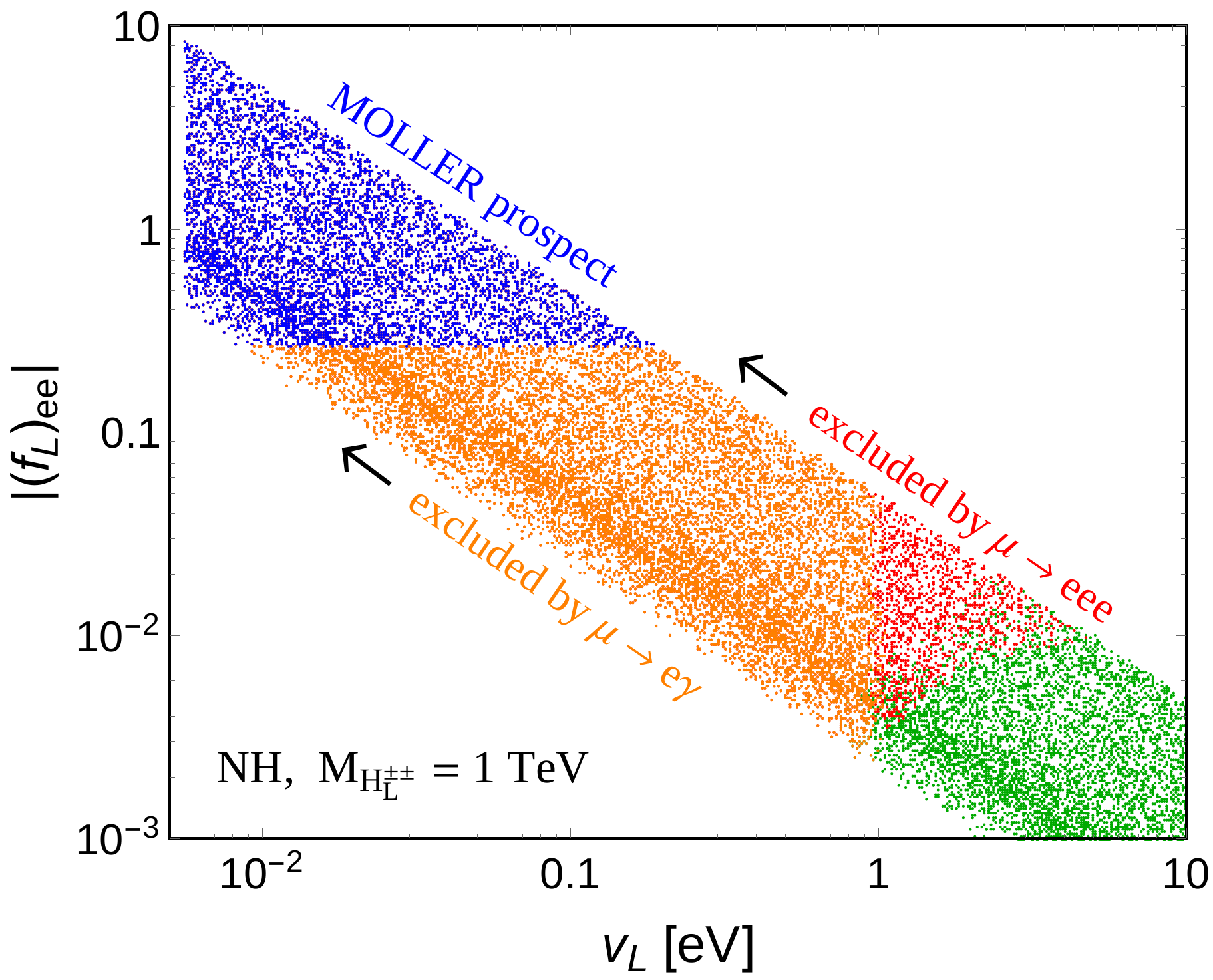}
\hspace{0.2cm}
  \includegraphics[height=0.32\textwidth]{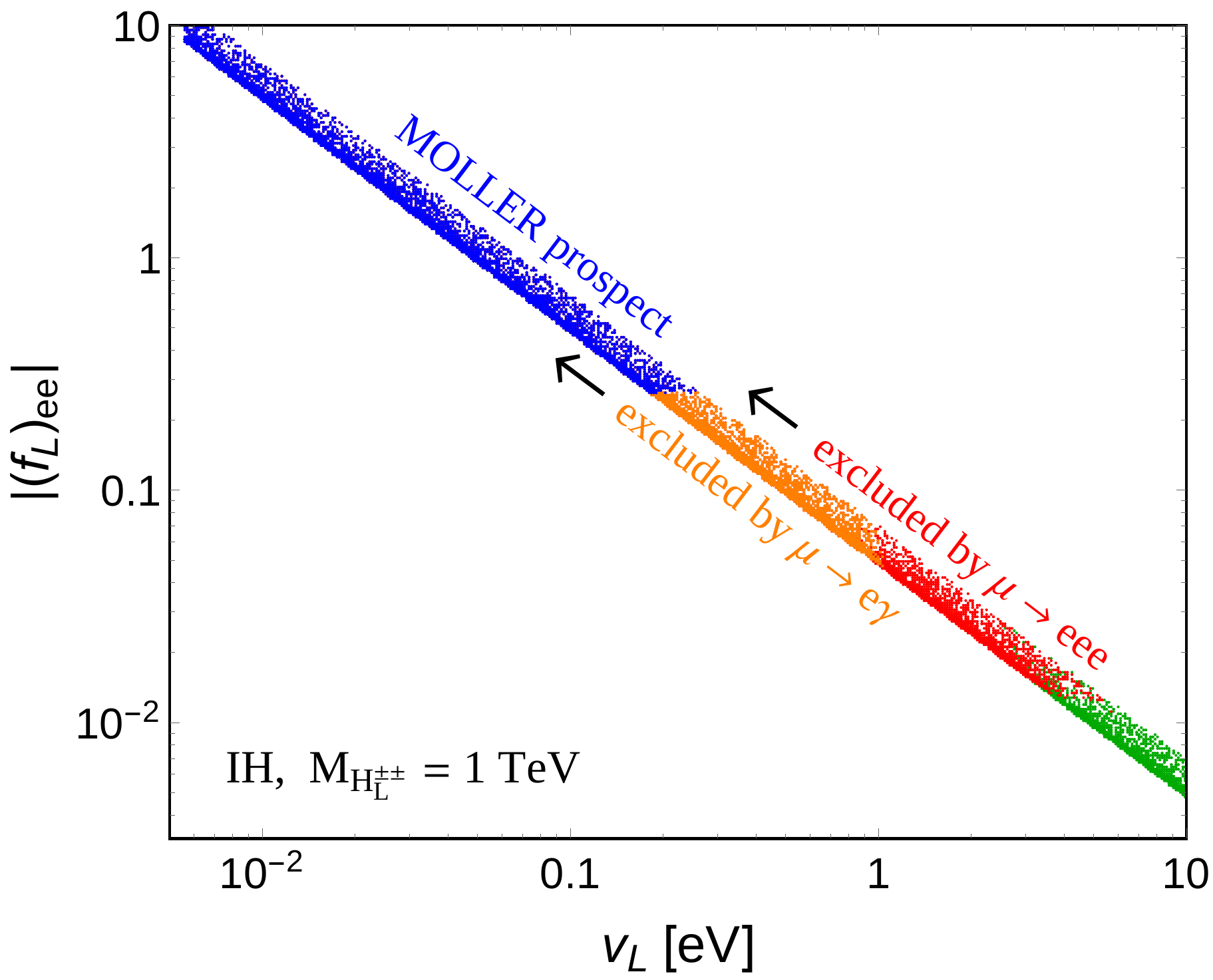}
  \caption{Scatter plots of the vev $v_L$ and the Yukawa coupling $|(f_L)_{ee}|$ in the type-II seesaw with $M_{H_L^{\pm\pm}} = 1$ TeV and for NH (left) and IH (right).  The blue points can be tested at MOLLER (Fig.~\ref{fig:left1}), while the orange points are excluded by $\mu \to e\gamma$, including the blue region as indicated by the arrow, and the red ones are excluded by $\mu \to eee$ including also the orange and blue regions as indicated by the arrow. The green ones are still allowed by both $\mu \to eee$ and $\mu \to e\gamma$ limits.}
  \label{fig:left2}
\end{figure*}

The CLFV decays $\mu \to eee$ and $\mu \to e\gamma$ depend on different combinations of the Yukawa couplings $(f_L)_{\alpha\beta}$, as shown in Eqs.~(\ref{eqn:mu2eee}) and (\ref{eqn:mu2egamma}). Nevertheless, one could still compare the results from searches for these CLFV processes  to the MOLLER prospect in Eq.~(\ref{eqn:moller}) in a straightforward way, as all the Yukawa entries $(f_L)_{\alpha\beta}$ are correlated by the neutrino data in the framework of type-II seesaw, as shown in Eq.~(\ref{eqn:neutrino}). To this end, with the same sets of randomly-scattered neutrino input parameters as above, we evaluate the corresponding $\delta A_{\rm PV}$ in MOLLER experiment, ${\rm BR} (\mu \to eee)$ and ${\rm BR} (\mu \to e\gamma)$, following the formulae in Eqs.~(\ref{eqn:mu2eee}) and (\ref{eqn:mu2egamma}), as functions of  $v_L$ and $|(f_L)_{ee}|$.  The results for the NH and IH are presented in the left and right panels of Fig.~\ref{fig:left2}, respectively.
In both panels, the blue points could be tested with the MOLLER experiment (Fig.~\ref{fig:left1}), while the orange points (including the blue ones) are excluded by the current $\mu \to e\gamma$ constraint. The red points (including the blue and orange points) are excluded by the current $\mu \to eee$ limit. As implied by Eqs.~(\ref{eqn:moller}), ~(\ref{eqn:mu2eee}) and (\ref{eqn:mu2egamma}), an observable $\delta A_{PV}$ in the MOLLER experiment is clearly precluded by the current limits from the CLFV decays $\mu \to eee$ and $\mu \to e\gamma$, even after taking into consideration the neutrino parameter uncertainties in Table~\ref{tab:neutriodata} and the unknown lightest neutrino mass.

For the loop-induced decay $\mu \to e\gamma$, we have summed up in Eq.~(\ref{eqn:mu2egamma}) all contributions involving an intermediate electron, muon or tau lepton,  corresponding respectively to the terms $\ell = e,\,\mu,\,\tau$ in the numerator of Eq.~(\ref{eqn:mu2egamma}). If the light neutrino masses are of NH, then the electron loop contribution, which is proportional to $(f_L)_{ee} (f_L)_{e\mu}^\ast$, can be sensitive to the lightest neutrino mass $m_0$ and much smaller than the muon and tau lepton contributions. Thus, in the left panel of Fig.~\ref{fig:left2}, the boundary for the $\mu \to e\gamma$ limit in orange is almost vertical and independent of $(f_L)_{ee}$. In contrast, the $\mu \to eee$ limit in red depends both on $(f_L)_{ee}$ and the vev $v_L$, as implied by Eq.~(\ref{eqn:mu2eee}). In the IH case, the contribution of electron loop to $\mu \to e \gamma$ is important, thus in the right panel of Fig.~\ref{fig:left2} the $\mu \to e\gamma$ limit is somewhat sensitive to $(f_L)_{ee}$.

In light of these results, we deduce that an observation of non-vanishing $\delta A_{\rm PV}$ by the MOLLER experiment would imply that the simplest type-II seesaw has to be extended to accommodate the deviation (assuming no other BSM contributions to the asymmetry). We discuss one such possibility, namely, the LRSM, in the following two  sections.

\section{Right-handed doubly-charged scalar in the Left-Right Extension of Type-II Seesaw}
\label{sec:lrsm}

\begin{figure*}[t!]
  \centering
  \includegraphics[height=0.25\textwidth]{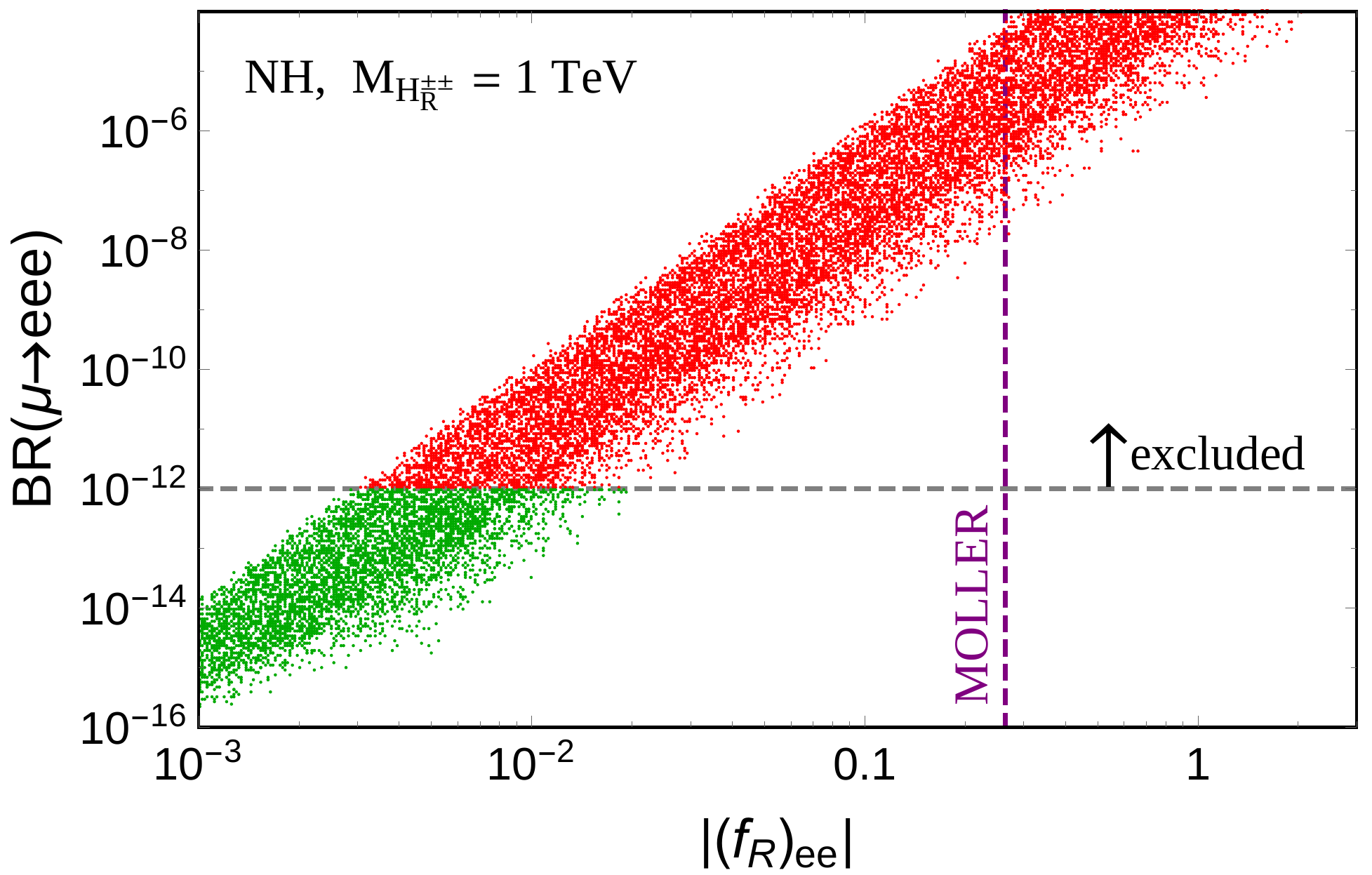}
  \includegraphics[height=0.25\textwidth]{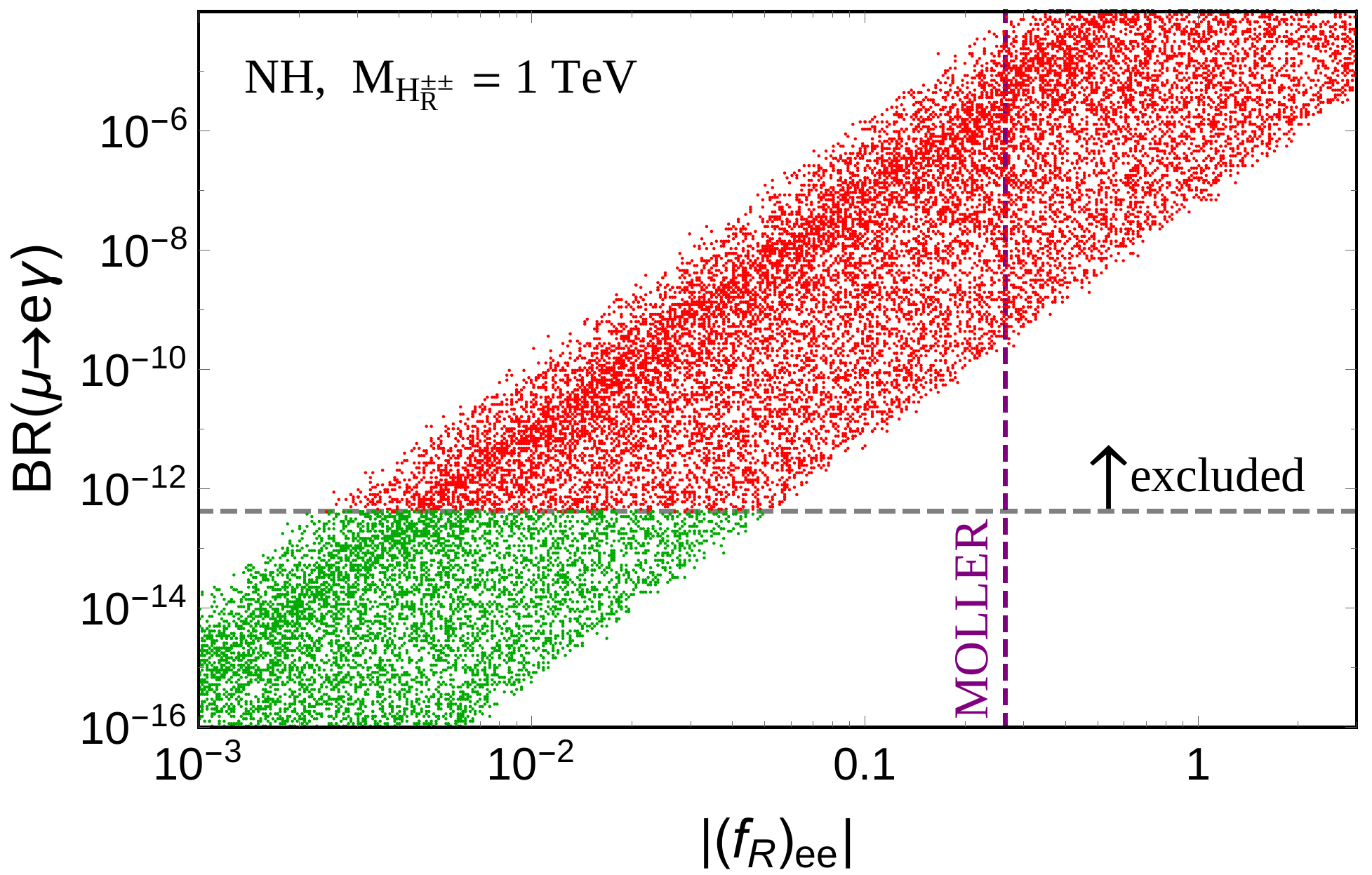} \\
  \includegraphics[height=0.25\textwidth]{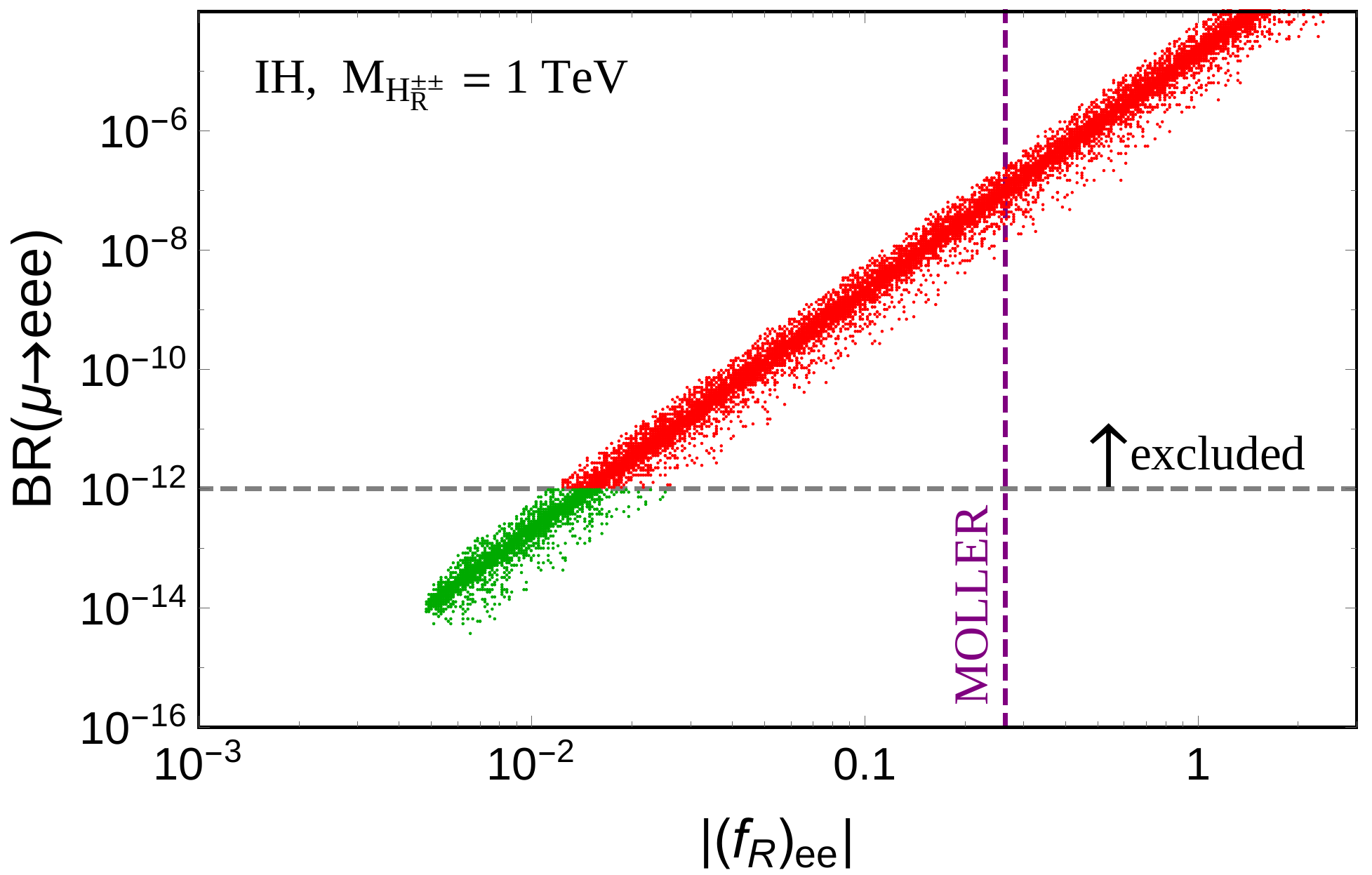}
  \includegraphics[height=0.25\textwidth]{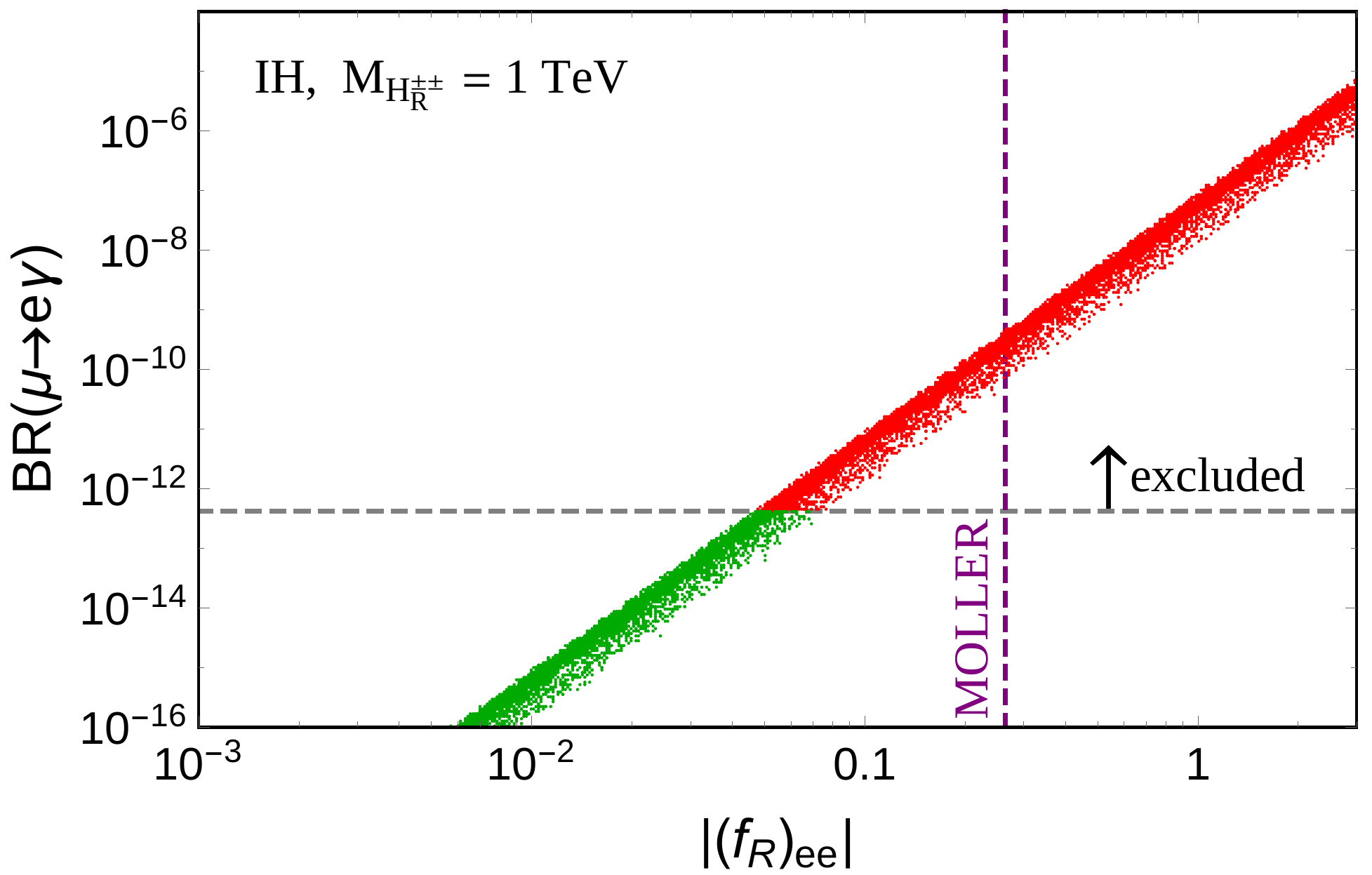}
  \caption{Scatter plots of ${\rm BR} (\mu \to eee)$ (left) and ${\rm BR} (\mu \to e\gamma)$ (right) as functions of $|(f_R)_{ee}|$ in the LRSM for NH (upper) and IH (lower), with the RH doubly-charged scalar mass $M_{H_R^{\pm\pm}} = 1$ TeV. The horizontal lines indicate the corresponding current CLFV limit (with the regions above these lines containing the red points excluded), and the vertical line gives the MOLLER prospect. }
  \label{fig:right}
\end{figure*}

The left-right symmetric model~\cite{LR1, LR2, LR3}, which provides a natural embedding of the type-II seesaw, contains two scalar triplets -- $\Delta_L$ and $\Delta_R$ -- that transform non-trivially under $SU(2)_L$ and $SU(2)_R$ respectively. In the limit of small mixing between all the neutral, singly-charged and doubly-charged scalars of $\Delta_L$ and $\Delta_R$, the LH triplet $\Delta_L$ can be identified as that in the type-II seesaw in Eq.~(\ref{eqn:DeltaL}); the RH triplet
\begin{align}
\label{eqn:DeltaR}
\Delta_R \ = \  \left(\begin{array}{cc}
\delta_R^+/\sqrt{2} & \delta_R^{++} \\
\delta_R^0 & -\delta_R^+/\sqrt{2}
\end{array}\right)
\end{align}
is the counterpart of $\Delta_L$ under parity and the RH doubly-charged scalar is identified as $H_R^{\pm\pm} = \delta_R^{\pm\pm}$. The triplet $\Delta_R$ couples to the RH lepton doublets $\psi_R =(N,\ell_R)^{\sf T}$ via the Yukawa interactions
\begin{eqnarray}
\label{eqn:lagrangian2}
{\cal L}_Y \ = \
- \left(f_R \right)_{\alpha\beta} \psi_{R,\,a\alpha}^{\sf T}C \varepsilon_{ab}
{\Delta}_R \psi_{R,\,b\beta} ~+~ {\rm H.c.},
\end{eqnarray}
with $N_\alpha$ the heavy RHNs. 
A non-zero vev of the neutral component $\langle \delta_{R}^0 \rangle = v_R/\sqrt2$ gives rise to the Majorana masses for the RHNs $M_N = \sqrt2 f_R v_R$, and parity symmetry dictates equality of the Yukawa couplings $f_R = f_L$. In the LRSM, the small neutrino masses receive, in principle, contributions from both type-I and type-II seesaw mechanisms:
\begin{eqnarray}
\label{eqn:neutrino2}
m_\nu \ \simeq \ - m_D M_N^{-1} m_D^{\sf T} + \sqrt2 f_L v_L  \,,
\end{eqnarray}
with $m_D$ the Dirac mass matrix. Since the focus of this study is on the type-II mechanism, we will assume here that the type-I seesaw contribution is small, {\it i.e.},   the LRSM is in the type-II dominance regime for neutrino mass generation (below, we comment briefly on the implications of relaxing this assumption). Since parity implies $f_L=f_R$ the heavy and light neutrino masses are related via $m_\nu / M_N \simeq v_L / v_R$. In this case, the RHN masses are proportional to those of the active neutrinos, rescaled by the vev ratio $v_R / v_L$, and the RHN mixing matrix $U_R$ is identical to the PMNS matrix $U$ in Eq.~(\ref{eqn:neutrino}).\footnote{A similar situation holds for the  the mixings in the quark sector~\cite{Zhang:2007da, Senjanovic:2014pva} under these assumptions.}

Importantly, all CLFV limits on the couplings and mass of LH doubly-charged scalar $H_L^{\pm\pm}$ -- such as the stringent constraints from $\mu \to eee$ and $\mu \to e\gamma$ --  also apply to the RH $H_R^{\pm\pm}$ sector; one needs only to replace the LH doubly-charged scalar mass and the Yukawa couplings in Eq.~(\ref{eqn:mu2eee}) and (\ref{eqn:mu2egamma}) by those for $H_R^{\pm\pm}$. Moreover,
the parity relation $f_R = f_L$ implies that the elements of the Yukawa coupling matrix $f_R$ are also related to the active neutrino masses and mixing angles, as in the minimal type-II seesaw. Using the same scan over neutrino mass and mixing parameters as in Section~\ref{sec:left}, we may compare the MOLLER reach with the limits from $\mu \to eee$ and $\mu \to e\gamma$: the plots of ${\rm BR} (\mu \to eee)$ and ${\rm BR} (\mu \to e\gamma)$ as functions of $|(f_R)_{ee}|$ are presented respectively in the left and right panels of Fig.~\ref{fig:right} with the RH doubly-charged scalar mass $M_{H_R^{\pm\pm}} = 1$ TeV, for both NH (upper) and IH (lower) cases. In these plots the horizontal lines are the current experimental LFV limits (with the regions above these lines excluded) and the vertical line indicates the MOLLER reach (Fig.~\ref{fig:prospect}). Again, the  parameter regions accessible to the MOLLER experiment have already  been excluded by the CLFV constraints for both NH and IH, irrespective of the neutrino parameter uncertainties.

In principle, both the LH and RH doubly-charged scalars can contribute to the LFV observables and the MOLLER asymmetry, and in this case their contributions will be added to each other. Parity symmetry implies they do so constructively, and their relative importance will depend on the magnitudes of the scalar masses. In the LRSM, if parity is violated, {\it i.e.} $f_L \neq f_R$ (see section \ref{sec:lrsm2} below), then for particular values of {phases of $f_{L,R}$, the LH and RH contributions may cancel against each other, thereby opening up an allowed window for MOLLER sensitivity. However, we do not entertain this fine-tuned possibility here.} 

\section{Right-handed doubly-charged scalar in LRSM with parity violation}
\label{sec:lrsm2}

\begin{figure*}[!t]
  \def\topdiff{0.25}
  \def\toppos{-1.5}
  \def\vertexstart{-4}
  \def\vertex{\vertexstart+1.5}
  \def\topoffset{0.75}

  \centering

  \noindent
  \begin{tikzpicture}[]
  \draw[quark,thick] (\vertexstart-1.5,1.2)node[left]{{\footnotesize$d$}} -- (\vertexstart-0.5,1.2);
  \draw[quark,thick] (\vertexstart-1.5,-1.2)node[left]{{\footnotesize$d$}} -- (\vertexstart-0.5,-1.2);
  \draw[quark,thick] (\vertexstart-0.5,1.2) --
  (\vertex,1.2)node[right]{{\footnotesize $u$}};
  \draw[quark,thick] (\vertexstart-0.5,-1.2) -- (\vertex,-1.2)node[right]{{\footnotesize$u$}};
  \draw[ZZ,thick](\vertexstart-0.5,1.2)--
  (\vertexstart,0.5)node[below=-3pt, left=4pt]{{\footnotesize$W$}} -- (\vertexstart+0.2,0.5);
  \draw[ZZ,thick](\vertexstart-0.5,-1.2)--
  (\vertexstart,-0.5)node[above=-3pt, left=4pt]{{\footnotesize$W$}} -- (\vertexstart+0.2,-0.5);
  \draw[thick] (\vertexstart+0.2,0.5)  --
  (\vertexstart+0.2,0)node[left]{{\footnotesize $\nu_i$}} -- (\vertexstart+0.2,-0.5);

  \draw[electron,thick] (\vertexstart+0.2,0.5) --
  (\vertex,0.5)node[right]{{\footnotesize $e^-$}};
  \draw[electron,thick] (\vertexstart+0.2,-0.5) -- (\vertex,-0.5)node[right]{{\footnotesize$e^-$}};
  \end{tikzpicture} \hspace{6pt}
  \begin{tikzpicture}[]
  \draw[quark,thick] (\vertexstart-1.5,1.2)node[left]{{\footnotesize$d$}} -- (\vertexstart-0.5,1.2);
  \draw[quark,thick] (\vertexstart-1.5,-1.2)node[left]{{\footnotesize$d$}} -- (\vertexstart-0.5,-1.2);
  \draw[quark,thick] (\vertexstart-0.5,1.2) --
  (\vertex,1.2)node[right]{{\footnotesize $u$}};
  \draw[quark,thick] (\vertexstart-0.5,-1.2) -- (\vertex,-1.2)node[right]{{\footnotesize$u$}};
  \draw[ZZ,thick](\vertexstart-0.5,1.2)--
  (\vertexstart,0.5)node[below=-3pt, left=4pt]{{\footnotesize$W_R$}} -- (\vertexstart+0.2,0.5);
  \draw[ZZ,thick](\vertexstart-0.5,-1.2)--
  (\vertexstart,-0.5)node[above=-3pt, left=4pt]{{\footnotesize$W_R$}} -- (\vertexstart+0.2,-0.5);
  \draw[thick] (\vertexstart+0.2,0.5)  --
  (\vertexstart+0.2,0)node[left]{{\footnotesize $N_i$}} -- (\vertexstart+0.2,-0.5);

  \draw[electron,thick] (\vertexstart+0.2,0.5) --
  (\vertex,0.5)node[right]{{\footnotesize $e^-$}};
  \draw[electron,thick] (\vertexstart+0.2,-0.5) -- (\vertex,-0.5)node[right]{{\footnotesize$e^-$}};
  \end{tikzpicture} \hspace{6pt}
  \begin{tikzpicture}[]
  \draw[quark,thick] (\vertexstart-1.5,1.2)node[left]{{\footnotesize$d$}} -- (\vertexstart-0.5,1.2);
  \draw[quark,thick] (\vertexstart-1.5,-1.2)node[left]{{\footnotesize$d$}} -- (\vertexstart-0.5,-1.2);
  \draw[quark,thick] (\vertexstart-0.5,1.2) --
  (\vertex,1.2)node[right]{{\footnotesize $u$}};
  \draw[quark,thick] (\vertexstart-0.5,-1.2) -- (\vertex,-1.2)node[right]{{\footnotesize$u$}};
  \draw[ZZ,thick](\vertexstart-0.5,1.2)--
  (\vertexstart,0.5)node[below=-3pt, left=4pt]{{\footnotesize$W_R$}} -- (\vertexstart+0.2,0);
  \draw[ZZ,thick](\vertexstart-0.5,-1.2)--
  (\vertexstart,-0.5)node[above=-3pt, left=4pt]{{\footnotesize$W_R$}} -- (\vertexstart+0.2,0);
  \draw[dashed,thick] (\vertexstart+0.2,0) --
  (\vertexstart+0.7,0)node[above]{{\footnotesize $H_R^{--}$}} -- (\vertexstart+1,0);

  \draw[electron,thick] (\vertexstart+1,0) --
  (\vertex,0.5)node[right]{{\footnotesize $e^-$}};
  \draw[electron,thick] (\vertexstart+1,0) -- (\vertex,-0.5)node[right]{{\footnotesize$e^-$}};
  \end{tikzpicture}
  \caption{Feynman diagrams for the parton-level $0\nu\beta\beta$ decays induced by the active neutrinos $\nu_i$ (left), the heavy RHNs $N_i$ (middle) and the RH doubly-charged scalar $H_R^{\pm\pm}$ (right), which correspond respectively to the $\eta_\nu$, $\eta_N$ and $\eta_{\delta_R}$ terms in Eq.~(\ref{eqn:0nubetabeta}).}
  \label{fig:diagram1}
\end{figure*}
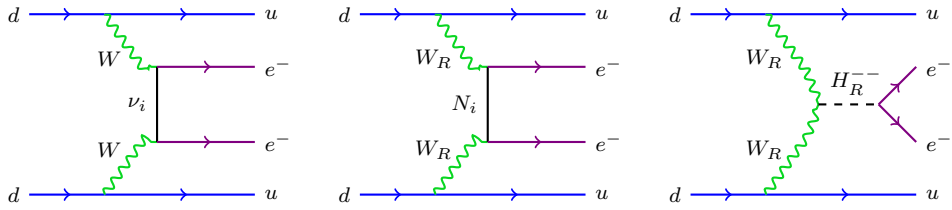

If parity is not completely restored in the LRSM at the TeV scale, the Yukawa couplings $f_{L,R}$ might not be equal. This possibility may also allow one to address some  theoretical issues for neutrino mass generation in the LRSM. Specifically,  the minimization conditions of the scalar potential require that $v_L \sim v_{\rm EW}^2/v_R$. This implies that, for TeV scale $v_R$, we have $v_L\sim {\cal O}$(GeV), which gives an unacceptably large type-II seesaw contribution  to the light neutrino masses $m_\nu \sim f_L v_L$ if $f_L = f_R \sim {\cal O}(1)$. One solution is to invoke significant cancellations between the type-I and type-II contributions [cf.~Eq.~(\ref{eqn:neutrino2})] to keep the neutrino masses at the sub-eV level. A more natural way is to eliminate the type-II seesaw contribution altogether: in a LRSM with $D$-parity breaking~\cite{CMP}, by introducing a parity-odd singlet with high-scale vev, one can give a large mass to $\Delta_L$ so that it decouples from the low-energy theory. Then the neutrino masses are generated via the type-I seesaw $m_\nu \simeq - m_D M_N^{-1} m_D^{\sf T}$.

\subsection{Neutrinoless double beta decay}

In the presence of parity-violation, one must consider an additional set of prospective constraints associated with $0\nu\beta\beta$ decays of nuclei, a lepton-number violating process that has not yet been observed. In general, if the light neutrinos are Majorana particles, then their exchange will induce $0\nu\beta\beta$-decay through the amplitude illustrated in the left panel of Fig.~\ref{fig:diagram1}.
In the LRSM, there are extra contributions from interactions of the heavy $W_R$,  $H_R^{--}$ boson and the RHNs $N_i$~\cite{Borah:2015ufa, Pritimita:2016fgr, Borgohain:2017akh, Ge:2015yqa, Dev:2013vxa, Dev:2014xea, Deppisch:2014zta, Mohapatra:1981pm, Hirsch:1996qw, Huang:2013kma}, corresponding to the right and middle panels, respectively, of Fig.~~\ref{fig:diagram1}. The $H_L^{--}$ contribution associated with the right panel of Fig.~\ref{fig:diagram1} is negligible, as it is suppressed by $(f_L)_{ee} v_L/M_{H_{L}^{\pm\pm}}^{2}$.
Neglecting the heavy-light neutrino mixing and the small $W - W_R$ mixing, the half lifetime of $0\nu\beta\beta$ can be factorized to be of the form~\cite{Barry:2013xxa}\footnote{If the $W - W_R$ mixing is sizable, then the nuclear matrix element (NME) for this contribution is enhanced by chiral symmetry, and in principle,  can compete with the light and heavy neutrino contributions in Eq.~(\ref{eqn:0nubetabeta}). See Ref.~\cite{Prezeau:2003xn} for more details.}
\begin{eqnarray}
\label{eqn:0nubetabeta}
\left[ T_{1/2} \right]^{-1} \ = \
G \, \left| {\cal M}_\nu \eta_\nu + {\cal M}_N \left( \eta_N + \eta_{\delta_R} \right) \right|^2 \,,
\end{eqnarray}
with $G$ the phase space factor,  ${\cal M}_\nu$ and ${\cal M}_N$  the NMEs for the diagrams with light and heavy neutrinos, respectively. The dimensionless factor $\eta_\nu = (m_\nu)_{ee} / m_e$ ($m_e$ being the electron mass) is the canonical term with the effective electron neutrino mass
\begin{eqnarray}
\label{eqn:mnuee}
(m_\nu)_{ee} & \ = \ &
\sum_i U_{ei}^2 m_{\nu,\,i}  \nonumber \\ & \ = \ &
m_1 c_{12}^2 c_{13}^2 +
m_2 s_{12}^2 c_{13}^2 e^{i\alpha} +
m_3 s_{13}^2 e^{i\beta}
\end{eqnarray}
encoding the Majorana phases $\alpha$ and $\beta$ {and the mixing angles $\theta_{ij}$ (with $c_{ij}\equiv \cos\theta_{ij}$, $s_{ij}\equiv \sin\theta_{ij}$).} If we have the LH doubly-charged scalar $H_L^{\pm\pm}$, its contribution $\eta_{\delta_L}$ is suppressed by the coupling to the SM $W$ boson (the small vev $v_L$), or effectively suppressed by the doubly-charged scalar mass i.e. $\eta_{\delta_L} \simeq
(m_e^2 / M_{H_{L}^{\pm\pm}}^{2}) \eta_\nu$~\cite{Chakrabortty:2012mh}, thus the contribution from $H_L^{\pm\pm}$ can be safely neglected.

\begin{figure*}[t!]
  \centering
  \includegraphics[width=0.4\textwidth]{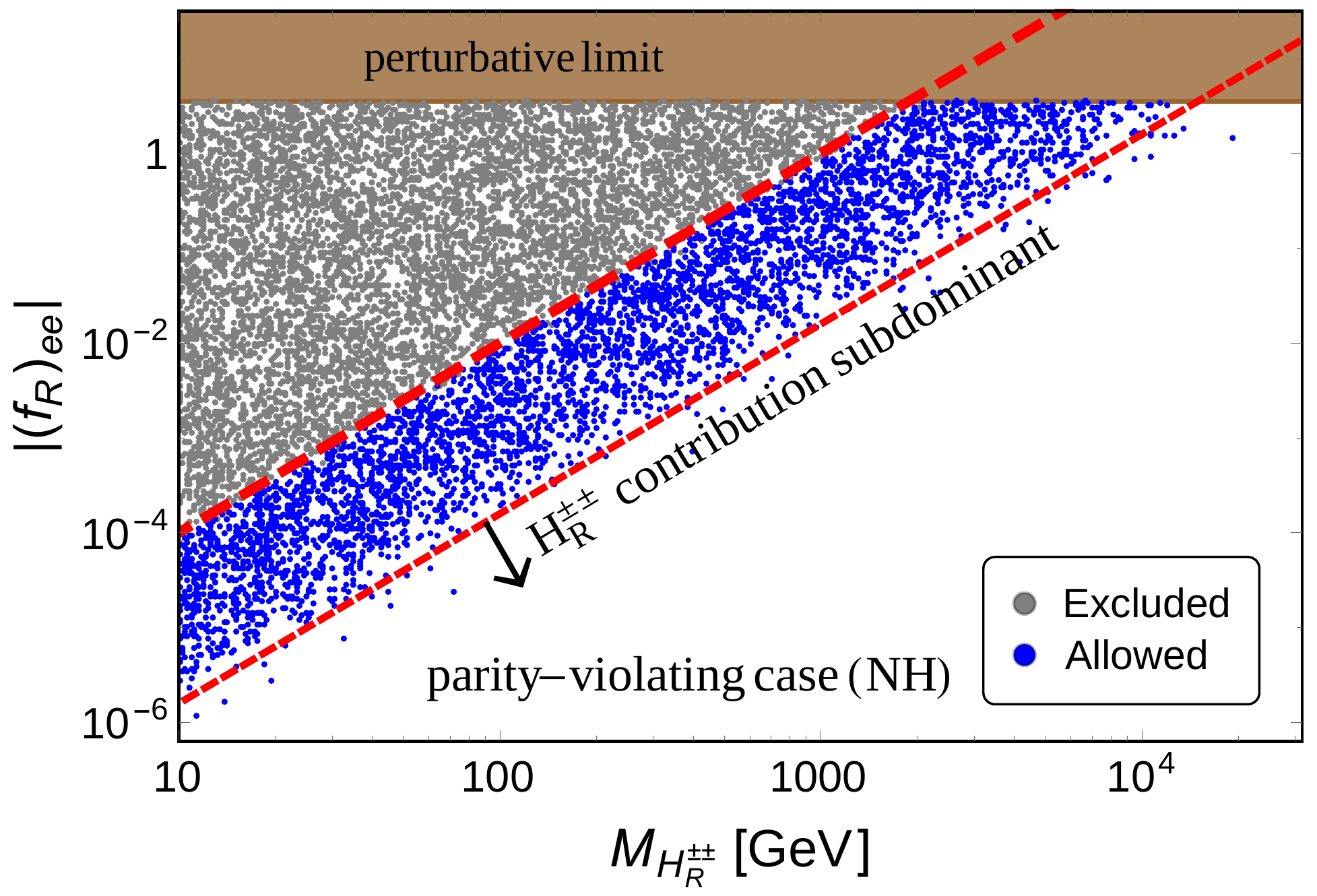}
  \includegraphics[width=0.41\textwidth]{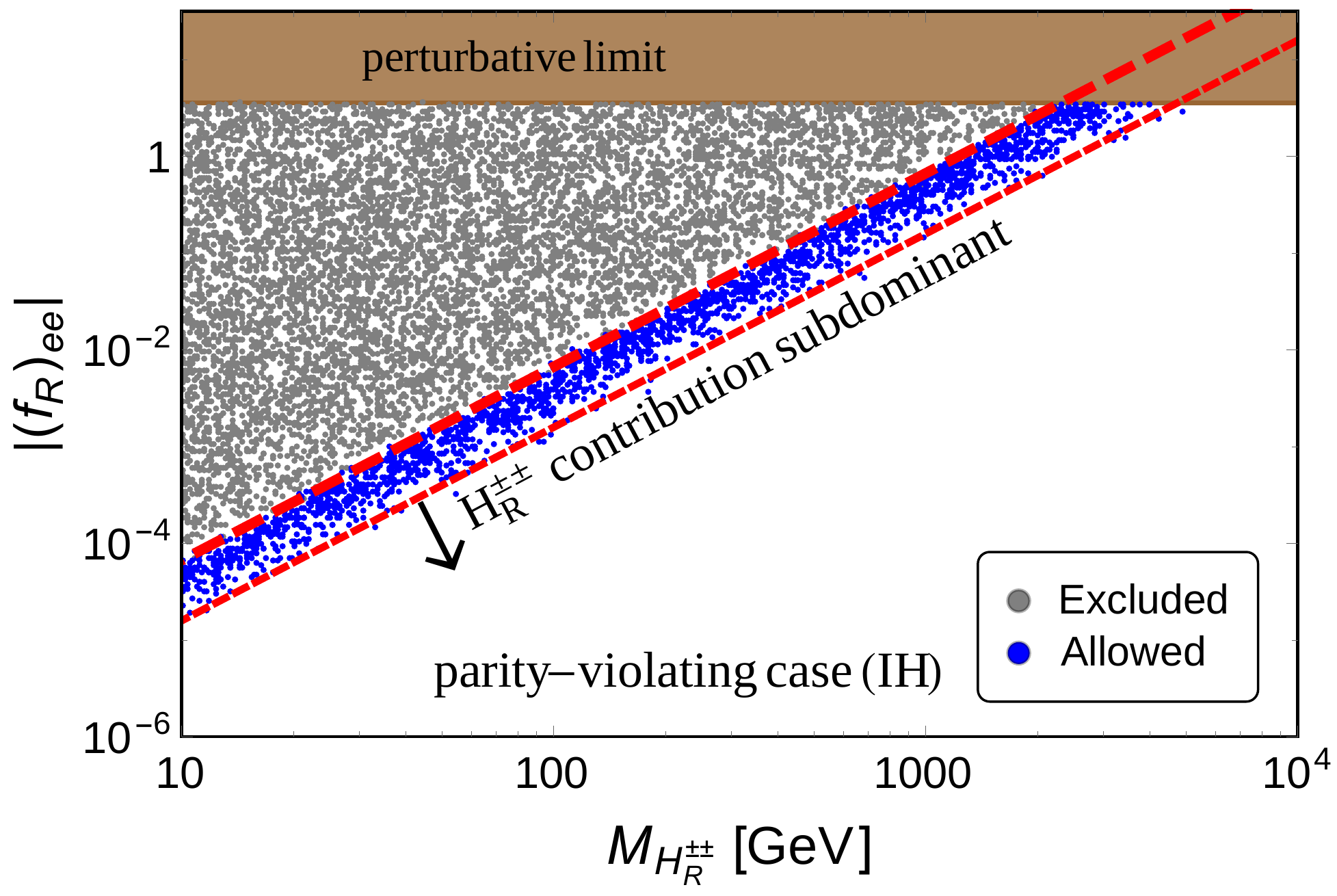}
  \caption{The scatter plots of $0\nu\beta\beta$ decay constraints on the RH doubly-charged scalar mass $M_{H_R^{\pm\pm}}$ and the Yukawa coupling $|(f_R)_{ee}|$ in the parity-violating LRSM with neutrino spectrum of NH (left) and IH (right) and with $v_R = 5\sqrt2$ TeV. All the gray points (regions above the long-dashed red line) are excluded by either KamLAND-Zen~\cite{KamLAND-Zen:2016pfg} or GERDA~\cite{Agostini:2018tnm} data, while those in blue are allowed. Below the short-dashed red line, the contribution of $H_R^{\pm\pm}$ to $0\nu\beta\beta$ decays is sub-dominant to the canonical light neutrino Majorana mass contribution. The brown region at top is excluded by the perturbativity requirement: $|(f_R)_{ee}| < \sqrt{4\pi}$.}
  \label{fig:0nubetabeta}
\end{figure*}

The last two terms in Eq.~(\ref{eqn:0nubetabeta}) are respectively from the RHN and $H_R^{\pm\pm}$ diagrams:
\begin{eqnarray}
\label{eqn:etaN}
\eta_N & \ = \ & m_p
\left( \frac{g_R}{g_L} \right)^4
\left( \frac{m_W}{M_{W_R}} \right)^4
\sum_i \frac{(U_R)_{ei}^2}{M_{N_i}}  \nonumber \\ & \ = \ &
\frac{m_p}{4}
\left( \frac{v_{\rm EW}}{v_{R}} \right)^4
\sum_i \frac{(U_R)_{ei}^2}{M_{N_i}} \,, \\
\label{eqn:etaDCSR}
\eta_{\delta_R} & \ = \ & m_p
\left( \frac{g_R}{g_L} \right)^4
\left( \frac{m_W}{M_{W_R}} \right)^4
\frac{ \sqrt2 (f_R)_{ee} v_R }{M_{H_R^{\pm\pm}}^2} \nonumber \\ &  \ = \ &
\frac{m_p}{2\sqrt2}
\left( \frac{v_{\rm EW}}{v_{R}} \right)^4
\frac{ (f_R)_{ee} v_R }{M_{H_R^{\pm\pm}}^2} \,,
\end{eqnarray}
where $m_p$ is the proton mass, $M_{N_i}$ the mass eigenvalues for the three heavy RHNs, and $U_R$ the RHN mixing matrix.  Note that there is essentially no dependence on the gauge coupling $g_R$ in Eq.~(\ref{eqn:etaDCSR}): 
at the amplitude level, the $W_R$ boson couples to the fermions or the scalar $H_R^{\pm\pm}$ with the strength $g_R^2$, which cancels out the $g_R$ dependence in the $W_R$ propagator. The RH doubly-charged scalar mass in Eq.~(\ref{eqn:etaDCSR}) is effectively suppressed by the $v_R$ scale.

The heavy and light neutrino contributions to the $0\nu\beta\beta$ decays have already been discussed on general grounds, {\it e.g.} in Refs.~\cite{Prezeau:2003xn, Cirigliano:2004tc, Dev:2013vxa, Dev:2014xea, Bambhaniya:2015ipg}. To set $0\nu\beta\beta$ decay limits on the RH doubly-charged scalar $H_R^{\pm\pm}$ and the coupling $(f_R)_{ee}$, we have to compare the three terms in Eq.~(\ref{eqn:0nubetabeta}) and identify the region in which the $H_R^{\pm\pm}$ contribution dominates. Let us first make the comparison of the last factors in Eq.~(\ref{eqn:etaN}) and (\ref{eqn:etaDCSR}), with $(f_R)_{ee} v_R \sim M_{N_i}$, we get the ratio $\eta_N / \eta_{\delta_R} \sim M_{H_R^{\pm\pm}}^2 / M_{N_i}^2$, which means that the doubly-charged scalar contribution is expected to be larger than that from the RHNs if $H_R^{\pm\pm}$ is lighter, i.e. $M_{H_R^{\pm\pm}} \lesssim M_{N_i}$.
If the RHNs $N_i$ are lighter than $H_R^{\pm\pm}$, then the contribution of $N_i$ to $0\nu\beta\beta$ (the middle diagram in Fig.~\ref{fig:diagram1}) is expected to be more important than that of $H_R^{\pm\pm}$ (the right diagram in Fig.~\ref{fig:diagram1}), which would weaken to some extent the $0\nu\beta\beta$ constraints on the doubly-charged scalar $H_R^{\pm\pm}$. In this sense, we are considering here the  scenario in which the $0\nu\beta\beta$ constraints are most likely to compete with those from other $f_R$-dependent observables.

Comparing then the $H_R^{\pm\pm}$ contribution in Eq.~(\ref{eqn:etaDCSR}) with the canonical term $\eta_\nu$ gives
\begin{eqnarray}
\label{eqn:ratio}
\frac{\eta_{\delta_R}}{\eta_\nu} \ = \ \frac{1}{2\sqrt2}
\left( \frac{v_{\rm EW}}{v_{R}} \right)^4
\left( \frac{(m_\nu)_{ee}}{(f_R)_{ee} v_R} \right)^{-1}
\left( \frac{m_e m_p}{M_{H_R^{\pm\pm}}^2} \right) \,. \nonumber \\
\end{eqnarray}
One can see that if the doubly-charged scalar mass $M_{H_R^{\pm\pm}} \sim {\rm TeV}$ and the Yukawa coupling $(f_R)_{ee} \sim {\cal O} (1)$, the contribution from $H_R^{\pm\pm}$ could be comparable to the $\eta_\nu$ term and thus get constrained by the limits from KamLAND-Zen~\cite{KamLAND-Zen:2016pfg} and  GERDA~\cite{Agostini:2018tnm}. The limits from EXO-200~\cite{Albert:2017owj}, CUORE~\cite{Alduino:2017ehq} and NEMO-3~\cite{Arnold:2018tmo} are somewhat weaker and are thus not explicitly considered here.

As in Section~\ref{sec:left} and \ref{sec:lrsm}, we scatter the neutrino data in Table~\ref{tab:neutriodata} within their $2\sigma$ ranges, the lightest neutrino mass $m_0 \in [0,\, 0.05 \, {\rm eV}]$, and adopt the NMEs
\begin{align}
{\cal M}_\nu&: & [2.58,\, 6.64] &\text{ for $^{76}$Ge, } &
[1.57,\, 3.85] &\text{ for $^{136}$Xe} \,, \nonumber \\
{\cal M}_N&: & [233,\, 412] &\text{ for $^{76}$Ge, } &
[164,\, 172] &\text{ for $^{136}$Xe} \,,
\end{align}
and the phase space factor $G = 5.77 \times 10^{-15}$ yr$^{-1}$ for $^{76}$Ge and $3.56 \times 10^{-14}$ yr$^{-1}$ for $^{136}$Xe from Ref.~\cite{Meroni:2012qf}. We set the RH scale $v_R = 5\sqrt2 $ TeV, and the results are shown in Fig.~\ref{fig:0nubetabeta} for both NH (left) and IH (right) cases. All the gray points (or the region above the long-dashed red line) are excluded by the current limits of $1.07 \times 10^{26}$ yrs for $^{136}$Xe from KamLAND-Zen~\cite{KamLAND-Zen:2016pfg} and $8.0 \times 10^{25}$ yrs for $^{76}$Ge from GERDA~\cite{Agostini:2018tnm}, at the 90\% CL, while the blue points are allowed. This implies an upper bound on $|(f_R)_{ee}|/ M_{H_R^{\pm\pm}}^2$, as shown by the solid brown line in Fig.~\ref{fig:final} for the IH (for NH, the bound is slightly weaker). Note that the dependence on the doubly-charged scalar mass and the Yukawa coupling is different from MOLLER sensitivity in Eq.~(\ref{eqn:moller}) and the CLFV limits in Eq.~(\ref{eqn:mu2eee}) and (\ref{eqn:mu2egamma}). For heavier $H_R^{\pm\pm}$ and/or smaller coupling $(f_R)_{ee}$, the contribution of $H_R^{\pm\pm}$ is suppressed [cf.~Eq.~(\ref{eqn:etaDCSR}) and (\ref{eqn:ratio})] and the $0\nu\beta\beta$ decays are dominated by the light neutrino diagrams [cf. the $\eta_\nu$ term in Eq.~(\ref{eqn:0nubetabeta})]. In such case, the KamLAND-Zen and  GERDA limits are no longer applicable to $H_R^{\pm\pm}$, which is indicated by  the short-dashed red line in Fig.~\ref{fig:0nubetabeta}.

\subsection{Collider constraints}
\label{sec:collider}

In the type-I dominance of LRSM, the neutrino data do not only depend on the coupling $f_R$ but also on the Dirac neutrino mass matrix $m_D$. As the matrix $m_D$ is completely unknown, we cannot constrain the $f_R$ couplings by solely using the neutrino data, and all the elements of $f_R$ can be considered as free parameters, though they are intimately connected to the heavy RHN masses through $M_N = \sqrt2 f_R v_R$. Moreover, most of the CLFV constraints such as those from $\mu \to eee$ and $\mu \to e\gamma$ also cannot be used to constrain the element $(f_R)_{ee}$, as they depend also on other entries of the $f_R$ matrix like $(f_R)_{e\mu}$ that -- in this scenario -- are not connected to $(f_R)_{ee}$ through neutrino properties. Thus, we consider other observables that depend directly on $(f_R)_{ee}$.

\begin{figure}[!t]
  \centering
  \includegraphics[width=0.4\textwidth]{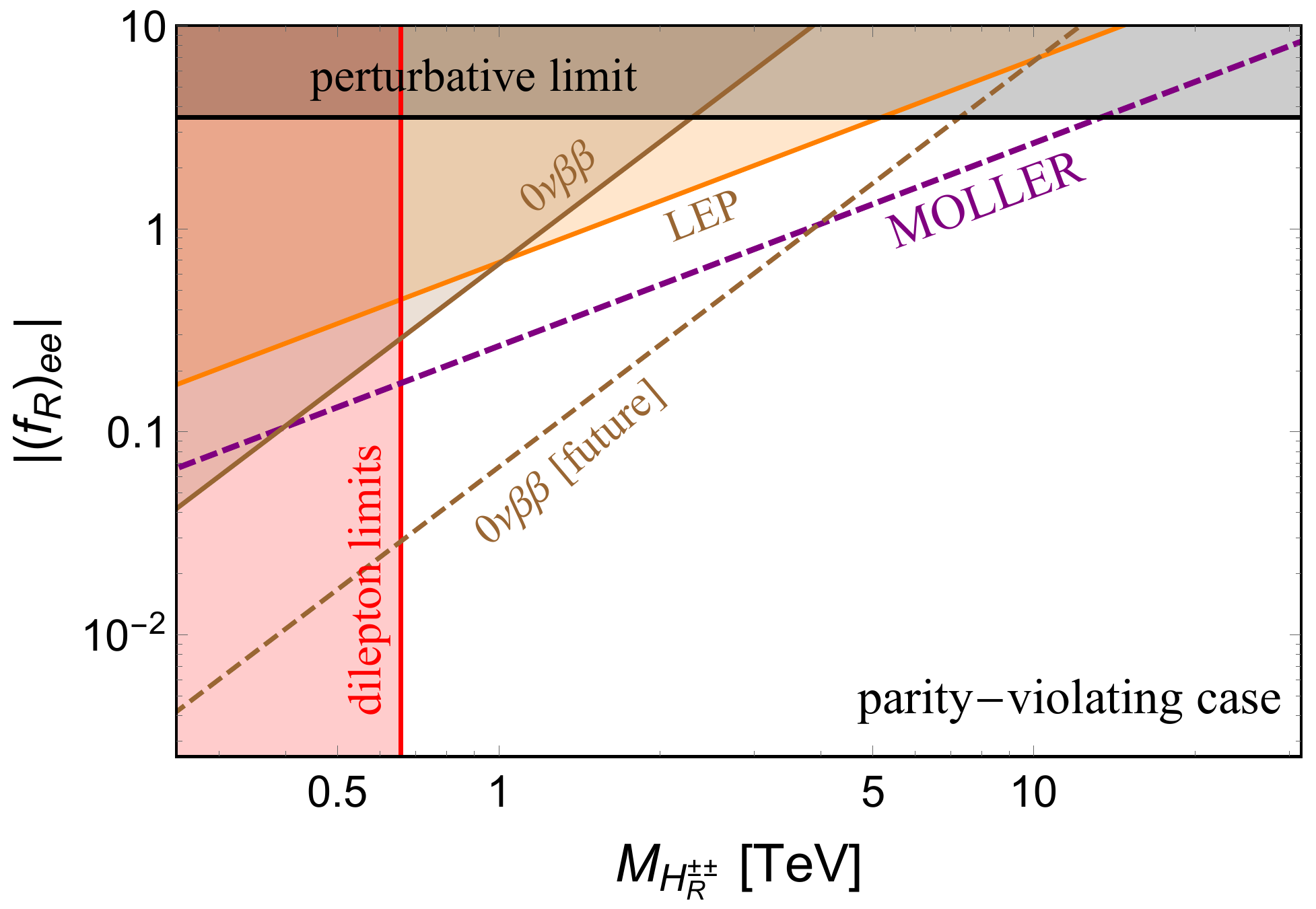}
  \caption{MOLLER prospect for the RH doubly-charged scalar mass $M_{H_R^{\pm\pm}}$ in the parity-violating LRSM and the coupling $|(f_R)_{ee}|$ (dashed purple line). We also show the same-sign dilepton limits from LHC 13 TeV assuming $H_R^{\pm\pm}$ decaying predominantly into electrons~\cite{Aaboud:2017qph, CMS:2017pet} (red), LEP $e^+e^- \to e^+e^-$ limit~\cite{Abbiendi:2003pr} (orange), and $0\nu\beta\beta$ limits from current KamLAND-Zen~\cite{KamLAND-Zen:2016pfg} and GERDA data~\cite{Agostini:2018tnm} (solid brown), as well as the future projection, both assuming an IH for the light neutrino spectrum. For the NH case, the $0\nu\beta\beta$ limit is slightly weaker (see Fig.~\ref{fig:0nubetabeta}). The dark gray region is excluded by the perturbativity limit $|(f_R)_{ee}| < \sqrt{4\pi}$.}
  \label{fig:final}
\end{figure}

The heavy $H_R^{\pm\pm}$ in the $t$-channel could mediate the Bhabha scattering $e^+ e^- \to e^+ e^-$ and interfere with the SM diagrams mediated by either $s$ or $t$-channel $\gamma$/$Z$. This alters both the total cross section and the differential distributions. If the Yukawa coupling $(f_R)_{ee}$ is of order one, $H_R^{\pm\pm}$ could be probed up to the TeV scale~\cite{Abbiendi:2003pr, Achard:2003mv}. By Fierz transformations, the coupling $(f_R)_{ee}$ of $H_R^{\pm\pm}$ contributes to the effective  four-fermion contact interaction
\begin{eqnarray}
\frac{1}{\Lambda_{\rm eff}^2}
(\bar{e}_R \gamma_\mu e_R)
(\bar{e}_R \gamma^\mu e_R) \,,
\end{eqnarray}
and is thus constrained by the LEP $e^+e^- \to e^+e^-$ data~\cite{Abdallah:2005ph} with $\Lambda_{\rm eff} \simeq M_{H_R^{\pm\pm}}/|(f_R)_{ee}|$ corresponding to the effective cutoff scale. It turns out the LEP data in Ref.~\cite{Abdallah:2005ph} set more stringent limits than those in Refs.~\cite{Abbiendi:2003pr, Achard:2003mv} and requires that $\Lambda_{\rm eff} \simeq M_{H_R^{\pm\pm}}/(f_R)_{ee} > 1.5$ TeV, somewhat weaker than the MOLLER sensitivity in Eq.~(\ref{eqn:moller}). The corresponding LEP  limit on the doubly-charged scalar mass $M_{H_R^{\pm\pm}}$ and the coupling $|(f_R)_{ee}|$ is shown in Fig.~\ref{fig:final} as the orange curve.

In the LRSM, the doubly-charged scalar $H_R^{\pm\pm}$ could decay into a pair of same-sign charged leptons $H_R^{\pm\pm} \to \ell_\alpha^\pm \ell_\beta^\pm$ or into a pair of (off-shell) heavy $W_R$ bosons $H_R^{\pm\pm} \to W_R^{\pm\, (\ast)} W_R^{\pm\, (\ast)}$ (note that the singly-charged component from $\Delta_R$ is eaten by the heavy $W_R$ boson after symmetry breaking)~\cite{Dev:2016dja}. The current $K$ and $B$ meson oscillation data require that the $W_R$ boson is heavier than roughly 3 TeV~\cite{Zhang:2007da, Bertolini:2014sua}; thus a TeV-scale (or lighter) doubly-charged scalar $H_R^{\pm\pm}$ decays predominantly  into same-sign dilepton pairs for a sizable Yukawa coupling $(f_R)_{\alpha\beta}$, and the most stringent dilepton limits are from the LHC 13 TeV data~\cite{Aaboud:2017qph, CMS:2017pet}. If $H_R^{\pm\pm}$ decays predominantly into $e^\pm e^\pm$ pairs, its mass is required to be larger than $657$ GeV, which is indicated by the vertical red line in Fig.~\ref{fig:final}. Note that the coupling of $H_R^{\pm\pm}$ to the SM $Z$ boson is proportional to $-2\sin^2\theta_W$, which leads to a destructive interference between the SM photon and $Z$-exchange amplitudes. On the other hand, in the case of $H_L^{\pm\pm}$, the coupling to the $Z$ boson is proportional to $1-2\sin^2\theta_W$, and the constructive interference of the SM photon and $Z$ diagrams renders the limits more stringent.

\subsection{Future collider prospects}

At a future high-energy lepton collider like CEPC~\cite{CEPC}, FCC-ee~\cite{Gomez-Ceballos:2013zzn}, ILC~\cite{Baer:2013cma} or CLIC~\cite{Battaglia:2004mw}, with an integrated luminosity of order 1 ab$^{-1}$, the coupling $(f_R)_{ee}$ could be probed to a much smaller value compared to LEP-II using the Bhabha scattering. Let us consider the CEPC 240 GeV as an explicit example. The cross section for Bhabha scattering $e^+ e^- \to e^+ e^-$ is about 1.4 times smaller than at LEP II. On the other hand, the  CEPC integrated luminosity is expected to be three orders of magnitude larger than that at LEP, where the integrated luminosity was merely 675 pb$^{-1}$~\cite{Abdallah:2005ph}. As a rough estimate, we rescale the LEP $e^+e^- \to e^+e^-$ limit in Ref~\cite{Abdallah:2005ph} by a factor of $[(\sigma_{\rm LEP}/\sigma_{\rm CEPC}) ({\cal L}_{\rm LEP}/{\cal L}_{\rm CEPC})]^{1/2}$, with $\sigma$ and ${\cal L}$ being the corresponding Bhabha cross section and integrated luminosity respectively. Given 1 ab$^{-1}$ of data, the prospective CEPC reach is 30 times stronger than that at LEP~\cite{Abdallah:2005ph}, as indicated by the dashed blue line in Fig.~\ref{fig:final2}.\footnote{If kinematically allowed, the doubly-charged scalar could also be singly produced, {\it e.g.} in the processes $e^+ e^- \to e^\pm e^\pm H_R^{\mp\mp}$ and $e^\pm \gamma \to e^\mp H_R^{\pm\pm}$; see Ref.~\cite{Dev:2018upe} for a complete analysis.}

At a future 100 TeV hadron collider~\cite{Arkani-Hamed:2015vfh} like SPPC~\cite{Tang:2015qga} or FCC-hh~\cite{fcc-hh} with a larger production cross section, the doubly-charged scalar $H_R^{\pm\pm}$ could be pair-produced in the Drell-Yan process and probed to a higher mass range than at LHC. Given an ultimate luminosity of 30 ab$^{-1}$, the $H_R^{\pm\pm}$ prospect could go up to 3.4 TeV in the Drell-Yan channel~\cite{Dev:2016dja}, with an RH scale of $v_R = 5\sqrt2$ TeV, as shown by the vertical dashed red line in Fig.~\ref{fig:final2}.

\subsection{MOLLER prospect}

\begin{figure}[!t]
  \centering
  \includegraphics[width=0.4\textwidth]{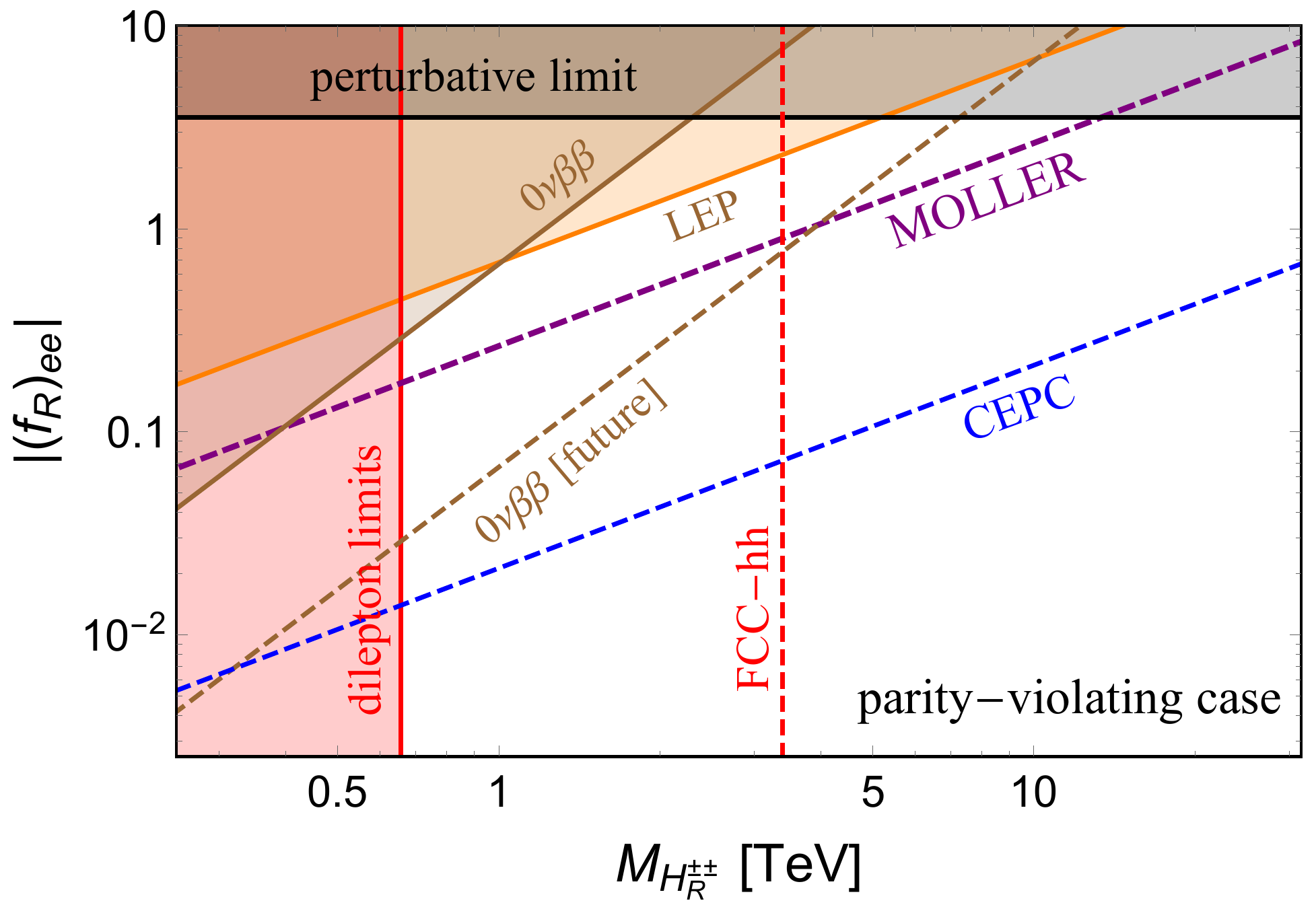}
  \caption{The same as in Fig.~\ref{fig:final}, along with the prospect of Bhabha scattering at CEPC 240 GeV with a luminosity of 1 ab$^{-1}$ (dashed blue), and the prospect at future 100 TeV collider with a luminosity of 30 ab$^{-1}$ (vertical dashed red line), assuming the RH scale $v_R = 5\sqrt2$ TeV. }
  \label{fig:final2}
\end{figure}

All the current limits on the doubly-charged scalar mass $M_{H_{R}^{\pm\pm}}$ and the coupling $|(f_R)_{ee})|$ are collected in Fig.~\ref{fig:final}, including those from the same-sign dilepton searches at LHC 13 TeV (red), the LEP $e^+e^- \to e^+e^-$ data (orange) and the $0\nu\beta\beta$ limit  (solid brown). The $0\nu\beta\beta$ limits correspond to the red long-dashed lines in Fig.~\ref{fig:0nubetabeta} for IH (for NH, the bound is slightly weaker and not shown here). All the shaded regions are excluded. The na\" ive perturbative limit $|(f_R)_{ee}| < \sqrt{4\pi}$ is indicated by the solid black line. We also indicate the prospective reach of future ton-scale $0\nu\beta\beta$ experiments~\cite{Agostini:2017jim}, assuming an increase in half-life sensitivity of two orders of magnitude (to $10^{28}$ years) compared with the present constraints. The representative future $e^+e^-$ and hadron collider reaches are indicated separately in Fig.~\ref{fig:final2} for comparison.


We highlight several salient features of these results:
\begin{itemize}
\item In contrast to the simplest type-II seesaw in Section~\ref{sec:left} and the RH $H_R^{\pm\pm}$ in parity-symmetric LRSM in Section~\ref{sec:lrsm}, there exists a considerable portion of the parameter space for $H_R^{\pm\pm}$ interactions in the parity-violating LRSM that could be tested by  MOLLER, as indicated by the dashed purple curves in Figs.~\ref{fig:final} and \ref{fig:final2}.
\item It is clear that for this scenario the high energy experiments are largely complementary to the low-energy fundamental symmetry tests. By direct production of $H_R^{\pm\pm}$, a high energy collider experiment could probe a lower doubly-charged scalar mass than MOLLER, but extend to much smaller values of the coupling $(f_R)_{ee}$. On the other hand, for a ${\cal O} (1)$ Yukawa coupling $(f_R)_{ee}$, the doubly-charged scalar mass $M_{H_R^{\pm\pm}}$ could be probed up to $\simeq 10$ TeV, which is far beyond the direct search capability of LHC or even future 100 TeV colliders.
\item The results of the MOLLER experiment could also have significant implications for the interpretation of $0\nu\beta\beta$ experiments. For an $H_R^{\pm\pm}$ mass above $\sim 5$ TeV,  one could anticipate a non-zero signal in the MOLLER experiment without a corresponding observable effect in the next generation $0\nu\beta\beta$ searches. On the other hand, for lighter masses, an observable $H_R^{\pm\pm}$ contribution to the MOLLER asymmetry would imply a non-zero signal in the future $0\nu\beta\beta$ experiments, barring any cancellation between the different amplitudes in Fig.~\ref{fig:diagram1}. Interestingly, the transition between these two mass regimes corresponds to the reach of a prospective future hadron collider. In short, the combination of these probes could help determine the mechanism of the  $0\nu\beta\beta$ process should a ton scale experiment yield a non-vanishing result.
\end{itemize}

\section{Conclusion}
\label{sec:conclusion}

Uncovering the dynamics responsible for generation of the non-vanishing light neutrino masses remains a forefront challenge for particle physics. In this work, we have studied how the interplay of various low-energy tests of fundamental symmetries with both neutrino oscillation phenomenology and present and future high-energy collider studies could probe the ingredients in the type-II seesaw mechanism and its extensions in left-right symmetric models. We have focused in particular on the impact of interactions mediated by the doubly-charged component of a complex scalar triplet, which is a key ingredient in these neutrino mass models.

For both the simplest type-II seesaw and its extension to a LRSM with parity symmetry (equality between the LH and RH triplet Yukawa couplings), searches for charged lepton flavor-violating processes such as $\mu\to eee$ and $\mu\to e\gamma$ provide the most powerful constraints.  In these scenarios, the flavor non-diagonal couplings are linked to the flavor-diagonal couplings by virtue of the neutrino mass matrix and, in the case of the parity-symmetric LRSM, by the assumption of parity symmetry. Combined with these relations, the present neutrino oscillation results and null results for CLFV searches imply that interactions mediated by the doubly-charged scalar would be too feeble to generate an observable effect in the next generation parity-violating
M{\o}ller experiment planned by the MOLLER collaboration.

On the other hand, when parity symmetry is broken in the LRSM at a scale much higher than the $SU(2)_R$-breaking scale, the connections with CLFV observables via neutrino phenomenology are lost. In this case, the MOLLER reach will exceed that of the present LEP II constraints and for sufficiently large Yukawa couplings $(f_R)_{ee}$ and doubly-charged scalar mass $M_{H_R^{\pm\pm}}$, the direct search limits from the LHC as well as from a future 100 TeV hadron collider. We also find that the results of the MOLLER experiment could have interesting implications for the interpretation of future ton-scale $0\nu\beta\beta$ experiments. Only with the advent of a future high luminosity $e^+e^-$ collider, such as the CEPC or FCC-ee, would the reach of high energy collider probes exceed that of the MOLLER experiment for this scenario.

\section*{Acknowledgments}

We acknowledge useful discussions and correspondence with Krishna Kumar. BD also thanks Albert de Roeck for a helpful comment and the organizers of WHEPP XIV at IIT, Kanpur and MASS2018 at CP$^3$-Origins, Odense, for the local hospitality, where part of this work was done.  YZ is grateful to the University of Maryland, College Park, for the hospitality and local support where part of the work was done. MJRM was supported in part under U.S. Department of Energy contract DE-SC0011095.

\end{document}